\documentclass[a4paper,fleqn]{cas-sc}

\usepackage{float}
\usepackage{graphicx}
\usepackage{caption}
\usepackage{subcaption}
\usepackage{adjustbox}
\usepackage[round,authoryear]{natbib}

\begin{document}
\let\WriteBookmarks\relax
\def\floatpagepagefraction{1}
\def\textpagefraction{.001}
\shorttitle{Riser Responses in Field Conditions Based on Gaussian Mixture Model}
\shortauthors{}

\title [mode = title]{Analysis of Full-scale Riser Responses in Field Conditions Based on Gaussian Mixture Model}             

\address[1]{SINTEF Ocean, Otto Nielsens veg 10, 7052 Trondheim, Norway}
\address[2]{SINTEF Digital, Forskningsveien 1, 0373 Oslo, Norway}
\address[3]{NTNU, Otto Nielsens veg 10, 7052 Trondheim, Norway}

\author[1]{Jie Wu}
\author[2]{Sølve Eidnes}
\author[1]{Jingzhe Jin}[type=editor,
                        auid=000,bioid=1,
                        orcid=0000-0003-0435-3084]
\cormark[1]

\author[1]{Halvor Lie}
\author[1]{Decao Yin}
\author[1]{Elizabeth Passano}
\author[3]{Svein Sævik}
\author[2]{Signe Riemer-Sørensen}


\cortext[cor1]{Corresponding author\\
        \indent \,\,\,\,\,\,  E-mail address: Jingzhe.Jin@sintef.no}

\begin{abstract}
Offshore slender marine structures experience complex and combined load conditions from waves, current and vessel motions that may result in both wave frequency and vortex shedding response patterns. Field measurements often consist of records of environmental conditions and riser responses, typically with 30-minute intervals. These data can be represented in a high-dimensional parameter space. However, it is difficult to visualize and understand the structural responses, as they are affected by many of these parameters. It becomes easier to identify trends and key parameters if the measurements with the same characteristics can be grouped together. Cluster analysis is an unsupervised learning method, which groups the data based on their relative distance, density of the data space, intervals, or statistical distributions. In the present study, a Gaussian mixture model guided by domain knowledge has been applied to analyze field measurements. Using the 242 measurement events of the Helland-Hansen riser, it is demonstrated that riser responses can be grouped into 12 clusters by the identification of key environmental parameters. This results in an improved understanding of complex structure responses. Furthermore, the cluster results are valuable for evaluating the riser response prediction accuracy.
\end{abstract}


\begin{highlights}

\item {The Helland-Hansen riser field measurement data was analyzed and grouped into different clusters by the identification of key environmental parameters representing wave, vessel motions and current conditions, using a Gaussian mixture model.}

\item {Improved understanding of marine riser responses due to complex and combined environmental loads has been obtained by evaluating characteristics of identified clusters.}

\item {Conditions for riser responses dominated by vortex induced vibrations (VIV) were identified and influences of environmental conditions on VIV were investigated.}

\item {Cluster analysis is a valuable tool to reduce the parameter dimensions so that the underlying physics can be better understood. In this study it contributed to the evaluation of the time-domain riser response prediction accuracy under various load conditions.}

\end{highlights}


\maketitle

\section{Introduction}
Slender marine structures, such as marine risers or power cables, are exposed to complex load conditions, i.e., wave, current, vessel motions and vortex induced vibrations (VIV) due to current. Among these, VIV in particular is often a critical design consideration as it can lead to fast accumulation of fatigue damage and amplified drag loads in face of strong currents. Extensive research has been performed over the years on this topic. However, most research efforts have been based on laboratory tests with a simplified structure model and flow conditions, and compromising scaling effects. There is also lack of a model test that can combine different load processes simultaneously as experienced in the field environmental conditions due to cost and physical limitations of the test setup. On the other hand, field measurements become increasingly available with the increased use of sensors. However, field measurement data have larger uncertainty compared to laboratory tests due to the relatively low sensor density and measurement quality for the complex response and load conditions.

Prediction accuracy using Frequency Domain (FD) semi-empirical prediction programs, such as VIVANA-FD \citep{Larsen2009}, Shear7 \citep{shear7} and VIVA \citep{viva99} are continuously improved through comparison against laboratory test data \citep{voie2017, vivbp2016}. However, there exist larger uncertainties in VIV response predictions when comparing with field measurement data, see e.g. \citep{Tognarelli2008, Ge2014}. In \citep{Tognarelli2008} it was concluded that "For full scale drilling risers without VIV suppression, data show that state-of-the-art analysis methods are, on average, inherently 30X (30 times) conservative on a maximum fatigue damage basis. This average bias may be reduced by adjusting the maximum lift in the lift curve utilized in the method; however, the ability to do this is limited due to the significant scatter in measured fatigue damage due to VIV." It is therefore important to reduce such uncertainties and increase safety in structure design against VIV loads. There is also a need to obtain insights on the complex physical processes experienced by the structure under field conditions. Some questions to be addressed are: 1) When will significant VIV responses occur? 2) How will different load conditions affect the structural responses? 3) How will the findings affect present prediction practices? 

Field measurements often consist of records of environmental conditions and structural responses typically with 30-minute intervals. Many environmental conditions will affect the structural responses, e.g., current profiles (speed, directionality and shearedness), direction and magnitude of wave force, and vessel motion induced loads. These parameters form a high-dimensional parameter space, which is difficult to visualize and interpret. Traditionally, the kurtosis value of the signal has been used as an indication in the case of a lock-in event. Kurtosis is defined as $m_{4}/m_{2}^2$, where $m_{4}$ and $m_{2}$ are the 4th and 2nd moments of the signal’s distribution, respectively. For a Gaussian process like the typical multi-frequency random vibrations, the kurtosis value is 3.0. For a sinusoidal process like the typical single mode lock-in event, the kurtosis value is equal to 1.5. However, this parameter alone provides limited information as multiple loads are normally present at the same time and the corresponding kurtosis will be between  1.5 - 3.0. 

It is easier to identify trends and key parameters if measurements with similar characteristics can be grouped together instead of examining cases individually. Cluster analysis is the process of grouping together data by common or similar characteristics \citep{Estivill2002, Xu2015, Gan2021}. The applications of cluster analysis are wide and across many different fields. Oftentimes clustering is an early key step within a larger framework. In data mining however, the identification of the different groups is the main purpose of doing cluster analysis. In some applications clustering is mainly used to identify similarities and differences between the individual data points. In the field of machine learning, cluster analysis is a subclass of unsupervised learning, since it is not dependent on training data assigned to predefined groups. Depending on the specific choice of algorithm, the data can be grouped based on different measures of distance, density of the data space, graphs, or statistical distributions. Based on the results of a previous study, on experimental data \citep{wu2020improved}, density and distribution based methods were applied for the field data in the present study.

It is known that the underlying mathematical models and empirical hydrodynamic coefficients of the present frequency domain VIV prediction programs are based on simplified assumptions. VIV responses of a slender beam are influenced by both the structural properties and flow characteristics. It was demonstrated in a concept study that the adaptive empirical hydrodynamic parameters may be applied to reduce the prediction discrepancy \citep{wu2020improved}. A semi-empirical time domain VIV model based on the synchronization concept was proposed by \citep{MT2014, MT2016, MT2017}. The pure cross-flow (CF) load model has been implemented in the prediction software VIVANA-TD after a systematic validation based on various laboratory test data in both constant flow \citep{Wu2020a} and oscillatory flow conditions \citep{Wu2022}. The model has been updated to describe CF and in-line (IL) coupled motions \citep{JV2019, Kim2021OE} and validated with respect to various deep water riser tests in 2D uniform and shear flow \citep{JV2018, JV2019, Kim2021OE, Kim2021JOMAE}. Reasonable VIV response prediction in 3D flow conditions were also obtained by this model \citep{KIM2022103057}. It can potentially also account for multiple simultaneously acting load processes \citep{wu2022b, yin2022a}. 

The objective of the present study is to investigate whether cluster analysis can help to 1) gain good insights of the physical process of different types of riser responses; 2) cluster field measurement data based on environmental load conditions. Gaussian mixture model (GMM) guided by domain knowledge has been applied to analyze the field measurements based on data of a drilling riser from the Helland-Hansen field in 1998 \citep{HH_report1998}. The cluster results are also used to explain the reasons of the riser response prediction discrepancy.

\section{Riser system and field measurement data}
\subsection{Riser system and instrumentation setup}
\begin{figure}[!h]
 		\centering
         \includegraphics[width=.8\textwidth]{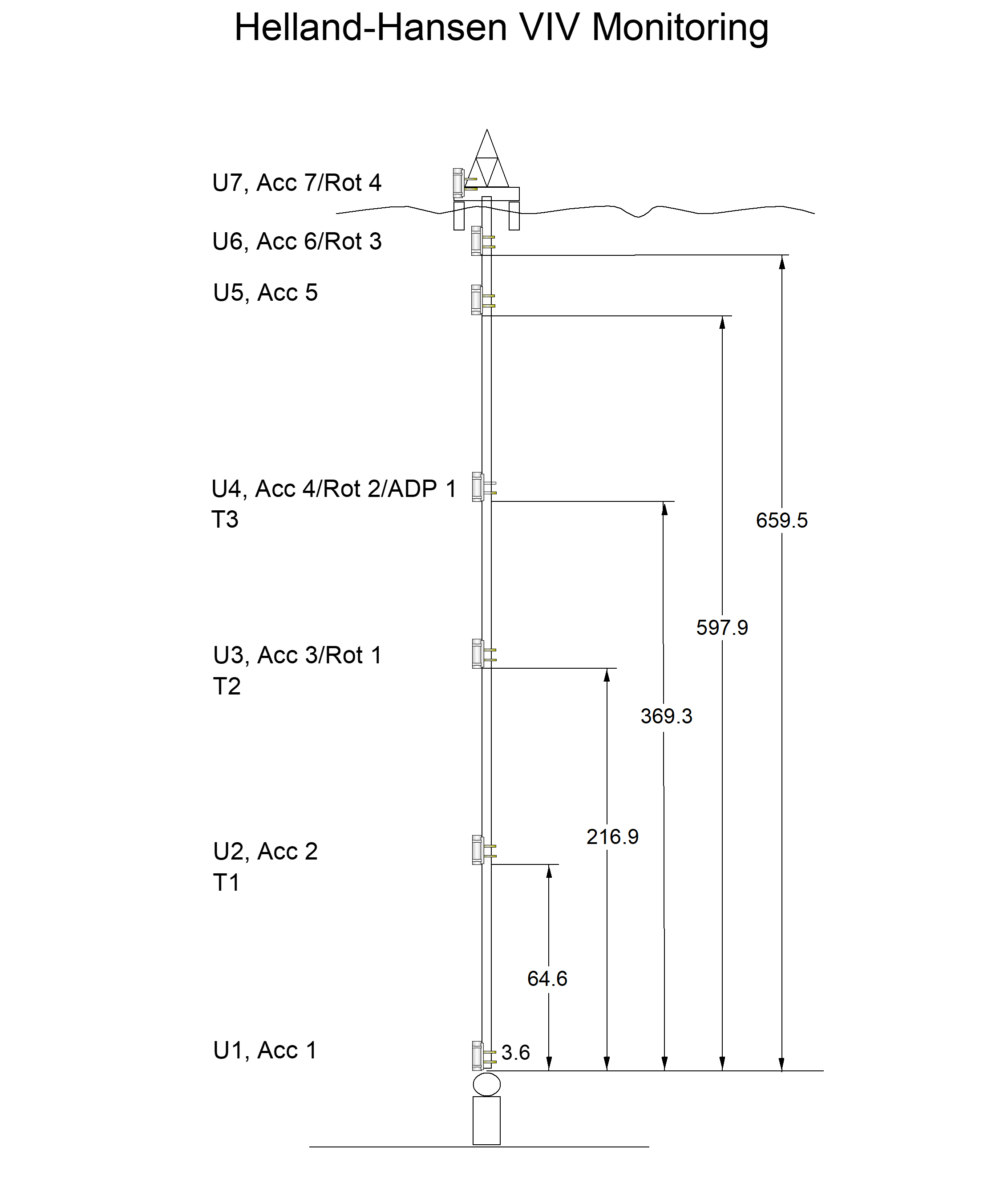}
         \caption{The Helland-Hansen riser system \citep{kaasen2000norwegian}, depth given in (m) }
         \label{fig:HH_riser_config}
\end{figure}

The Helland-Hansen drilling riser is 688 m long and the diameter of the bare riser section and the buoyancy element is 0.5 m and 1.13 m respectively. Most parts of the riser are covered by the buoyancy elements. The riser system is shown in Fig.\ \ref{fig:HH_riser_config} and the properties of the riser are summarized in Table \ref{tab:Riser_data}. The eigenfrequency of the riser system has been estimated and is shown in Table \ref{table:Riser_eigenfrq}; the eigenfrequency can vary due to the changes in weight and tension during the drilling operations. The tension was not measured during the measurement campaign, but it was estimated to be about 3000 kN to 4000 kN. The riser and vessel motions were measured as well as the current. A total of 916 sets of records were collected, each of 34 minutes in length \citep{HH_report1998}.

\begin{table}[!h]
\centering
\caption{Helland-Hansen riser properties}
\begin{tabular}{|l|l|} 
\hline
\textbf{Parameter}                             & \textbf{Value}    \\ 
\hline
Riser length (m)~                             & 682.75    \\ 
\hline
Submerged length (m)                          & 670.46    \\ 
\hline
Length of part with buoyancy at lower end (m) & 34        \\ 
\hline
Length of part with buoyancy at upperend (m)  & 100       \\ 
\hline
Total length with buoyancy (m)                & 503       \\ 
\hline
Outer diameter, bare riser (m)                & 0.5334    \\ 
\hline
Inner diameter, bare riser (m)                & 0.5017    \\ 
\hline
Outer diameter of riser + buoyancy (m)        & 1.13      \\ 
\hline
Mass per length unit, bare incl. mud (kg/m)   & 720.53    \\ 
\hline
Mud density (kg/m\textsuperscript{3})         & 1420      \\ 
\hline
Axial stiffness (N)                           & 5.42E+09  \\ 
\hline
Bending stiffness (Nm\textsuperscript{2})                         & 1.82E+08  \\
\hline
\end{tabular}
\label{tab:Riser_data}
\end{table}

\begin{table}[!h]
\centering
\caption{Eigenfrequency of the Helland-Hansen riser system \citep{kaasen2000norwegian}}
\begin{tabular}{|l|l|} 
\hline
\textbf{Eigenfrequency} & \textbf{Value (Hz)}  \\ 
\hline
f\textsubscript{1}     & 0.022 - 0.031   \\ 
\hline
f\textsubscript{2}     & 0.047 - 0.056   \\ 
\hline
f\textsubscript{3}     & 0.071 - 0.079   \\ 
\hline
f\textsubscript{4}     & 0.1 - 0.11      \\ 
\hline
f\textsubscript{5}     & 0.12 - 0.13     \\
\hline
\end{tabular}
\label{table:Riser_eigenfrq}
\end{table}

\begin{figure}[!h]
 		\centering
         \includegraphics[width=.7\textwidth]{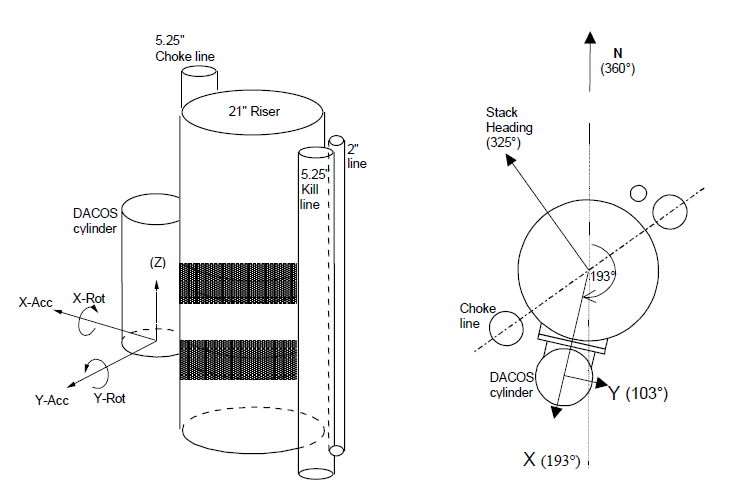}
         \caption{Orientation of measurement on the riser \citep{kaasen2000norwegian}.}
         \label{fig:Riser_orientation}
\end{figure}

The instrument system to measure riser responses on Helland-Hansen consisted of six instrument containers attached to the riser in positions shown on Fig.\ \ref{fig:HH_riser_config}. Each instrument unit contained motion sensors, data acquisition hardware and batteries. The main sensors were accelerometers for measurement of horizontal acceleration in two orthogonal axes, X and Y, as shown in Fig.\ \ref{fig:Riser_orientation}. In three of the instrument containers the accelerometers were supplemented with sensors for measuring angular velocity about the X- and Y-axes. 

A seventh instrument container was installed in the drilling vessel, including both accelerometers and angular velocity measurement. 
Current was measured by a number of acoustic doppler current profilers (ADCP), and a rotor-type current meter measured current near the seafloor.  The current speed and direction were measured at a number of depths. Current measurement for the top 100 metres of the water column was excluded due to the disturbance from the drilling vessel’s thrusters.  

\newpage
\subsection{Field measurement data processing}
The field measurement data has been analyzed in earlier studies \citep{HH_report1998, HH_NDPreport1999,kaasen2000norwegian}, but additional analysis was required in this study, as summarised in this section.

242 sets of the measured data have been analyzed and used. The riser experienced multiple environmental loads simultaneously, i.e., wave loads, VIV loads, vessel motions, etc. Low frequency vessel motion appears at 0.02 Hz, which is also close to the eigenfrequency of the first mode. The response at wave load frequency was typically broad-banded, which is between 0.08 - 0.2 Hz. VIV can occur at single or multiple frequencies depending on the flow speed and its profile. The highest expected VIV frequency is in the range of the wave frequency.   

\subsubsection{Current data analysis}
The measured current data has been processed to identify a main current direction. The shearedness and directionality of the current were defined and calculated.

\textbf{Main current direction:} 
The riser motion is measured in a coordinate system X-Y which is defined by the orientation of the instrument cylinders, see Fig.\  \ref{fig:Riser_orientation}. A new coordinate system $X_{C}$-$Y_{C}$ is found by a counter-clockwise rotation about the vertical axis. The axis $X_{C}$ is defined as the principal or main current direction and axis $Y_{C}$ is normal to the principal direction. 

\textbf{Directionality:} The horizontal current ($U_{Y_{C}}$) that is not in the main direction ($U_{X_{C}}$) is given by a spread parameter,  \textit{sprcoeff},  defined as follows:

\begin{equation}
\textit{sprcoeff}=\frac{\text{RMS}(U_{Y_{C}})}{\text{RMS}(U_{X_{C}})}=\sqrt{\frac{\sum_{i=1}^{n} (U^2_{Y_{C,i})})}{\sum_{i=1}^{n} (U^2_{X_{C,i})})}}
\end{equation}

It is assumed that the shedding process is insensitive to the sign of the velocity. For a pure 2-D current profile,  \textit{sprcoeff} becomes zero. For a case with RMS($U_{Y_{C}}$) equal to RMS($U_{X_{C}}$),  \textit{sprcoeff} becomes unity.

\textbf{Shearedness:} The shear of a current profile is defined as the ratio between the velocity variation $\triangle U$ and the average velocity $\overline{U}_{X_{C}}$. In many of the observed current profiles, current velocity varied a lot with a rather random pattern and a more general shear parameter is therefore defined:

\begin{equation}
\textit{shcoeff}=\frac{\sigma(U_{X_{C}})}{\overline{U}_{X_{C}}}=\frac{\sqrt{1/n\sum_{i=1}^{n} (|{U_{X_{C,i}}}|-\overline{U}_{X_{C,i}})^2}}{1/n\sum_{i=1}^{n} |U_{X_{C,i}}|)}
\end{equation}

The current profile tends to be uni-directional with increasing speed and the shearedness of the profile also increases. 

\subsubsection{VIV response analysis}
Using a modal approach, the lateral displacements along the length of the riser were reconstructed from the gravity-contaminated acceleration measurements, and the response was calculated as a weighted sum of a set of eigenfunctions. 

The VIV response frequency ($f_v=S_t U/D$) can be estimated based on the equivalent Strouhal number ($S_t$) of an oscillating cylinder \citep{Potts2018} and the current speed ($U$). Large scatter on Strouhal number is observed as it is affected by many parameters, e.g., the Reynolds number, roughness and model test setups. Strouhal number values around 0.25 - 0.3 can be expected for the Reynolds number range $1-7 \times 10^5$ based on the maximum current speed and the external diameter of 1.13 m. This means that the expected VIV response frequency ($f_v$) is about 0.025 Hz when the current speed is 0.1 m/s. This is close to the first eigenfrequency (0.022 - 0.031 Hz) of the riser and the first mode may be excited, without considering other factors. E.g., the actual dominating frequency may deviate from the Strouhal frequency when the flow speeds vary strongly along the length of the riser. The highest estimated VIV load frequency is in the range of wave frequencies.

\subsubsection{Vessel motion estimation}
The drilling vessel subject to wave, current and wind loads moves around its equilibrium position above the wellhead. Such motions can occur both at wave frequency (around 0.1 Hz) and at much lower frequency around 0.02 Hz. The low frequency vessel motion also introduces relative flow speeds at the riser top end. This speed can be the same magnitude as the current speed in some cases. Direct integration of acceleration measurement at vessel is sensitive to errors at such a low frequency. Therefore, the vessel motion was estimated by the modal approach using all of the acceleration measurements on the riser.
The measured vessel acceleration represents the resulted magnitude and direction of the wave loads. 

\subsubsection{Data used in the present study}\label{selected_data}

All of the data has been transformed to the coordinate system defined by the main current direction. The data can be classified into two categories, environmental load parameters and riser response parameters, as shown in Table  \ref{table:selected_data}. The environmental load parameters were applied as inputs to the clustering algorithm, while the riser response parameters were used to evaluate the cluster results. It is expected that riser responses will have the same characteristics if it is subject to similar load combinations, provided that the key load processes are correctly represented by these parameters.\\

\begin{table}[!h]
\centering
\caption{Selected data in present study}\label{table:selected_data}
\begin{tabular}{|l|l|l|} 
\hline
\multicolumn{2}{|l|}{\textbf{ Parameters~}}                 & \textbf{Comment}                                                                                                                                                              \\ 
\hline
\multirow{7}{*}{\textbf{Environmental load }} & Umax~       & Maximum current                                                                                                                                                               \\ 
\cline{2-3}
                                              & shcoeff~    & Current shearedness                                                                                                                                                           \\ 
\cline{2-3}
                                              & sprcoeff~   & Current directionality                                                                                                                                                        \\ 
\cline{2-3}
                                              & TopVeloX~   & \begin{tabular}[c]{@{}l@{}}Vessel velocity root mean square (rms) value \\in the main current direction\end{tabular}                                                          \\ 
\cline{2-3}
                                              & TopVeloY~   & \begin{tabular}[c]{@{}l@{}}Vessel velocity root mean square (rms) value \\normal to the main current direction\end{tabular}                                                   \\ 
\cline{2-3}
                                              & TopAccX~    & \begin{tabular}[c]{@{}l@{}}Vessel acceleration root mean square (rms) value \\in the main current direction\end{tabular}                                                      \\ 
\cline{2-3}
                                              & TopAccY~    & \begin{tabular}[c]{@{}l@{}}Vessel acceleration root mean square (rms) value \\normal to the main current direction\end{tabular}                                               \\ 
\hline
\multirow{4}{*}{\textbf{Riser response }}     & Freq\_dom~  & \begin{tabular}[c]{@{}l@{}}Dominating frequency of estimated riser displacement \\in the main current direction based on the modal analysis\end{tabular}                      \\ 
\cline{2-3}
                                              & Ydisp\_max~ & \begin{tabular}[c]{@{}l@{}}Maximum displacement root mean square (rms) value along the riser \\normal to the main current direction based on the modal analysis\end{tabular}  \\ 
\cline{2-3}
                                              & AccX3/4~    & \begin{tabular}[c]{@{}l@{}}Acceleration root mean square (rms) value at sensor no. 3/4 \\in the main current direction\end{tabular}                                           \\ 
\cline{2-3}
                                              & AccY3/4~    & \begin{tabular}[c]{@{}l@{}}Acceleration root mean square (rms) value at sensor no. 3/4 \\normal to the main current direction\end{tabular}                                    \\
\hline
\end{tabular}
\end{table}

\newpage
\section{Cluster method}
\subsection{Choice of clustering algorithms}\label{Clustering algorithms}
The many clustering algorithms from the literature can be divided into different categories. Among these categories there are partitioning methods, hierarchical clustering, spectral clustering, density-based clustering and distribution-based clustering.

\textit{Partitioning methods} use a distance-based metric to cluster the points based on their similarity. The clusters obtained are spherical-shaped in the parameter space, and not overlapping. K-means clustering belongs to this category \citep{lloyd1982least}. When the data consists of a known number of clearly separated clusters, this method works well. For the Helland-Hansen data, however, partitioning methods were deemed not effective as we do not expect clearly seperated clusters.

\textit{Hierarchical methods} form clusters by evaluating the maximum distance needed to connect parts of the cluster \citep{ward1963hierarchical, press2007numerical}, so that different clusters are formed at different distances. Thus, the algorithm does not provide a single partitioning of the data set, but instead provide a hierarchy of clusters. Hierarchical methods can be further divided into two types: agglomerative and divisive. \textit{Spectral clustering} uses the spectrum of the similarity matrix of the data to reduce the number of dimensions before clustering in the low-dimension space \citep{von2007tutorial}. Based on the results of using these of the experimental data considered in \citep{wu2020improved}, neither of these categories are deemed promising in our case.

In \textit{density-based methods}, HDBSCAN is an example method where clusters are defined as areas of higher density of data points in the parameter space \citep{kriegel2011density}. Different clusters are separated by areas of low density, and data points in such sparse areas are classified as noise. Unlike partitioning methods and hierarchical clustering, density-based clustering can capture arbitrarily shaped clusters.

Gaussian mixture is a form of \textit{distribution-based clustering} \citep{attias2000variational, press2007numerical}. It is a probabilistic model that assumes that the data points originate from a given number of overlapping Gaussian distributions with unknown locations and sizes. Thus the number of clusters must be given as an input parameter, and no data points will be classified as noise. Since the Helland-Hansen data is from continuous distributions of physical origin and has strong overlap in a high dimensional parameter space, and can thus be reasonably approximated by overlapping Gaussian distributions, this algorithm is chosen in the present study. 

\subsection{Clustering with Gaussian mixture models}
The Gaussian mixture models used here are based on the seminal papers by Banfield and Raftery \citep{Banfield1993model} and Celeux and Govaert \citep{Celeux1995gaussian}. The description of the method given in the following is mostly based on the latter reference. In GMM, it is assumed that the data $D = \left\{x_1, x_2, \ldots, x_n\right\}$, with independent observations $x_i \in \mathbb{R}^d$ for $i=1,\dots,n$, comes from a random vector with density
\begin{equation}
f(x) = \sum_{k=1}^K p_k \Phi(x \, \lvert \,  \mu_k,  \Sigma_k),
\end{equation}
where $p_k>0$ are the mixing proportions such that $\sum_{k=1}^K p_k=1$, and $\Phi(x \, \lvert \,  \mu,  \Sigma)$ is the density of a Gaussian distribution with mean $\mu$ and variance $\Sigma$. Common assumptions on $\Sigma$ are given in \citep{Celeux1995gaussian}. Then the goal is to find the parameters $\theta =\left\{ p_1,\ldots, p_{K-1}, \mu_1, \ldots, \mu_K,\Sigma_1, \ldots, \Sigma_K \right\}$ that maximizes the log-likelihood
\begin{equation}
L(\theta \, \lvert \, x_1, \ldots, x_n) = \sum_{i=1}^n \log \left( \sum_{k=1}^K p_k\Phi(x_i \, \lvert \, \mu_k, \Sigma_k) \right).
\end{equation}
This optimization task is usually performed in GMM, as in other mixture models, by the expectation-maximization (EM) algorithm of Dempster et al.\ \citep{Dempster1977maximum}.
In an iterative procedure and starting with an initial guess $\theta^0$, the expectation (E) step at iteration $j$ of the algorithm consists of estimating the conditional probability $t_k(x_i)$ that $x_i$ comes from the $k$-th mixture component, given the estimated $\theta^j$. This probability is given by
\begin{equation}
t_k(x_i) = \frac{ \hat{p}_k \Phi(x_i, \hat{\mu}_k, \hat{\Sigma})}{ \sum_{l=1}^K \hat{p}_l \Phi(x_i, \hat{\mu}_l, \hat{\Sigma}_l)}.
\end{equation}
In the maximization (M) step, updated maximum likelihood estimates $\hat{p}_k, \hat{\mu}_k, \hat{\Sigma}_k$ are computed using $t_k(x_i)$ as conditional mixing weights, and this gives $\theta^{j+1}$.

Here, $k$-means \citep{MacQueen1967some, Hartigan1979algorithm} was applied to initialize the parameters $\theta^0$, as suggested in \citep{McLachlan2019finite}. 100 random initialization of the $k$-means algorithm were used to make the results more consistent. Then 100 iterations of the EM algorithm was performed for each $\theta^0$ obtained. Only the best result was kept, i.e.\ the parameters with the largest likelihood.

\subsection{Validation of cluster analysis}

Statistical methods for cluster evaluation are commonly divided into two categories: internal and external evaluation \citep{Halkidi2002cluster, Gan2021}. These terms reflect whether the test data are obtained internally or externally with respect to the clustering process. Sometimes the terms supervised and unsupervised evaluation are used, but it is important to not confuse supervised validation of clustering with supervised learning. In addition there is a third class of approaches, called relative evaluation, which is not based on statistical tests and can be preferable when computational cost is an issue \citep{Halkidi2002clustering}. This was not considered in the present study.

An internal evaluation typically results in a single quality score for the clustering in question. Several different methods with corresponding scores exist, including the Davies--Bouldin index, the Dunn index and the silhouette index \citep{Arbelaitz2013extensive}. The motivation behind each of these methods is the intuitive idea that an item in a given cluster should be more similar to the other items in that cluster than items in other clusters. On the other hand, an external evaluation approach can typically consist of comparing cluster results with an existing ground truth classification, manual evaluation by a human expert, or indirect evaluation by evaluating the utility of the clustering in its intended application \citep{Pfitzner2009characterization}.

Both internal and external evaluation suffer from fundamental limitations. Internal evaluation is effective for comparing different clusters against each other, but the scores obtained will be highly biased towards assumptions made in the clustering processes. Furthermore, it is effective for comparing the results with respect to the different optimization objectives used for clustering and evaluation, where you get a high score if they would yield similar results. Similarly, external evaluation is biased towards the ground truth used. This ground truth is not always reliable; if it was very reliable, unsupervised learning would probably not be the preferred method. Furthermore, if the clusters are evaluated by humans, this evaluation will be highly subjective and not necessarily consistent. With this in mind, the evaluation performed in this study is most useful for identifying bad clusters, more than giving an accurate measure of the quality or utility of the end results.

\subsubsection{Internal evaluation method}

The clusters obtained by GMM were evaluated using silhouette scores \citep{Rousseeuw1987silhouettes} in this study. There is no known efficient algorithm for performing clustering with the objective given by the silhouette index. However, it is one of the most used internal criteria for evaluation, and can be seen as a measure of how well-defined the clusters are. Specifically, a silhouette value is calculated for each item in the data set. It is based on: 1) the mean distance between an item and all other items in the same cluster; 2) the mean distance between an item and all other items in the next nearest cluster. Formally, given a distance metric $d : \mathbb{R}^d \times \mathbb{R}^d \rightarrow \mathbb{R}$, 1) is found by calculating
\begin{equation}
    a(i) = \frac{1}{\lvert A \rvert-1} \sum_{j\in A,j\neq i} d(i,j),
\end{equation}
for the item $i \in A$, where $\lvert A \rvert$ denotes the number of items in cluster $A$. Then 2) is given by
\begin{equation}
    b(i) = \min_{B \neq A} \frac{1}{\lvert B \rvert} \sum_{j\in B} d(i,j),
\end{equation}
where $b(i)$ is the smallest mean distance of $i$ to all points in any other cluster, of which $i$ is not a member. The cluster with this smallest mean dissimilarity is said to be the "neighboring cluster" of $i$ because it is the next best fit cluster for point $i$.

The silhouette value is given by
\begin{equation}
    s(i) = \frac{b(i)-a(i)}{\max \left\{a(i), b(i)\right\}}
\end{equation}
if $\lvert A \rvert > 1$ and $s(i)=0$ if $\lvert A \rvert = 1$. Thus $-1 \leq s(i) \leq 1$, and the closer the score is to one, the more well-defined the clustering is. Values near zero indicate overlapping clusters.

By calculating silhouette values for all data points, an average for the whole data set and averages for each of the clusters can be obtained. The former provides a measure of how well-defined the clustering is as a whole, and thus scores for comparing different initializations and assumptions, e.g. with respect to the number of clusters. The latter gives a measure of how well-defined individual clusters are, and can thus be used to the quantify the confidence of the different clusters.

\subsubsection{External evaluation method} 
As mentioned above, external evaluation also comes with some issues: if we have accurate "ground truth" labels, then clustering is not needed. However, accurate ground truth labels usually do not exist. In addition, pre-existing labels only reflect one possible partitioning of the data set, which does not imply that there does not exist a different, and maybe better, clustering.

\begin{figure}[!h]
 		\centering
         \includegraphics[width=.6\textwidth]{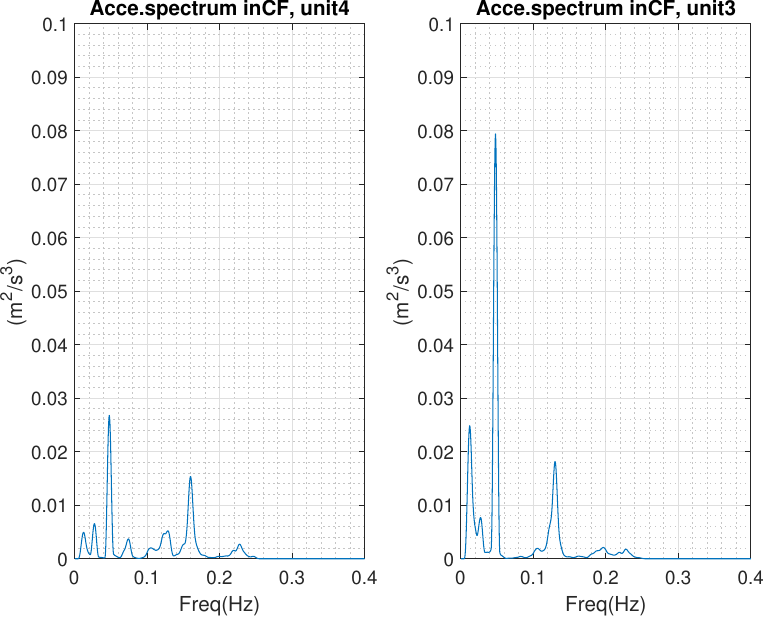}
         \caption{Example of measured acceleration spectrum at sensor unit 3 and 4 from 3rd of May 1998.}
         \label{fig:Acc_3_4_spec}
\end{figure}

Here, manual clustering was carried out by examining the frequency contents of the measured riser acceleration signals. 
Examples of acceleration spectra at both unit 3 and 4 in cross flow (CF) direction are shown in Fig.\ \ref{fig:Acc_3_4_spec}, from the measurement record taken on 3rd of May 1998 with maximum current velocity 0.2 m/s, current shearedness 0.37 and directionality parameter 0.79. For this case, the acceleration responses consist of responses due to VIV load at 0.05 Hz, wave load excitation at 0.1-0.2 Hz and low frequency motion excitation around 0.02 Hz.   

The maximum acceleration spectrum value at unit 3 or 4 in CF direction (normal to the main current direction) is used as reference quantity in the manual clustering. In the example case, the peak spectrum value is 0.079 $\mathrm{m}^2/\mathrm{s}^3$ at unit 3 and 0.027 $\mathrm{m}^2/\mathrm{s}^3$ at unit 4. The larger value of the two, at unit 3, is taken as reference quantity for this case. This is found efficient in finding those cases with responses dominated by VIV load, as the acceleration spectra often have high peaks and narrow bandwidth. On the other hand, the acceleration spectra for those cases dominated by wave load often have round peaks and wide bandwidth. For this reason, a relatively high level of acceleration spectrum value was applied to identify cases dominated by VIV load while a lower level was applied in identifying cases dominated by wave loads.  

\begin{figure}[!h]
 		\centering
         \includegraphics[width=.7\textwidth]{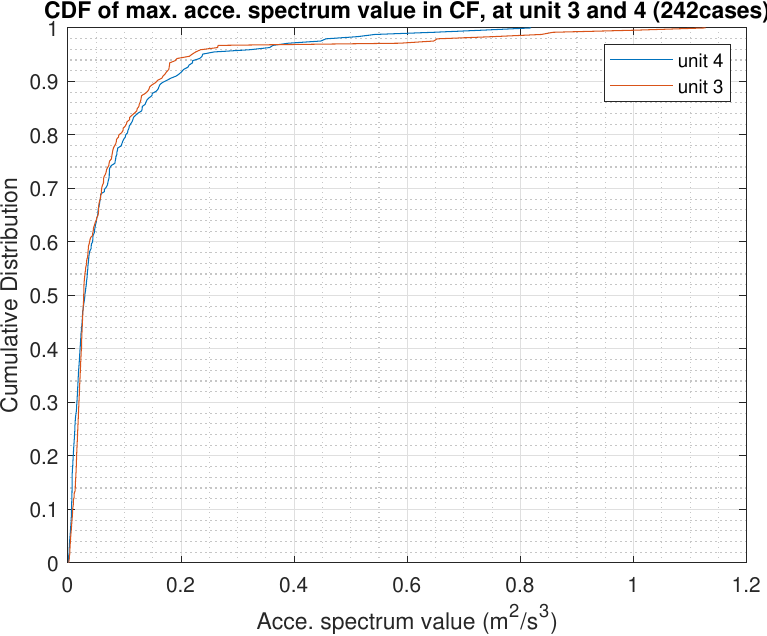}
         \caption{Cumulative distribution of maximum acceleration spectrum at sensor unit 3 and 4}
         \label{fig:CDF_acc_3_4}
\end{figure}

To obtain an overview of the maximum acceleration spectrum values for all 242 cases, the corresponding cumulative distribution functions (CDF) for both unit 3 and 4 are shown in Fig.\ \ref{fig:CDF_acc_3_4}. The CDFs of both units are in general close to each other. At CDF level of 0.7, the corresponding maximum acceleration spectrum value is 0.06 $\mathrm{m}^2/\mathrm{s}^3$ and at CDF level of 0.3, it is 0.02 $\mathrm{m}^2/\mathrm{s}^3$, for both units.  

The manual clustering was defined based on these two CFD levels: 0.7 for cases with relatively large responses; 0.3 for cases with relatively small responses; and 0.3-0.7 for cases with intermediate responses. Strouhal frequency was used to estimate the expected VIV response frequency. Based on these conditions, the response was manually classified as follows:

\begin{itemize}
\item Responses dominated by VIV load: spectrum peak value at either unit 3 or 4 larger than 0.06 $\mathrm{m}^2/\mathrm{s}^3$ and the corresponding frequency is close to the Strouhal frequency (difference < 0.04 Hz). 75 cases were identified as VIV load dominated cases.   

\item Small response cases: cases at both units smaller than 0.02 $\mathrm{m}^2/\mathrm{s}^3$. 61 cases were identified as small response cases.  

\item Responses dominated by wave load: for the remaining 107 cases, cases with frequency at maximum acceleration spectrum value at either unit 3 or 4 larger in wave frequency range and far away from Strouhal frequency (difference > 0.04 Hz) were recognized as wave load dominated cases. 52 cases were identified as wave load dominated cases.  

\item Responses subject to VIV, wave load and low frequency top motion:  The remaining 54 cases were observed to include combined load effects and the response magnitude was intermediate in general.  
\end{itemize}

The above criteria are based on the domain knowledge and can be subjective. The local measurements may not represent the global responses in some cases, e.g., the sensor happens to be located at a small response area, and the actual frequency may be lower than Strouhal frequency. In addition, there can be multiple loads acting simultaneously and multi-frequency response can be present even for VIV load dominant cases, as shown in Fig.\ \ref{fig:Acc_3_4_spec}. However, this method provided an independent evaluation based only on measured responses. Confidence on the cluster results is obtained when GMM using only environmental load parameters gives consistent clustering results compared to the manual evaluation. 

 242 evaluated cases were further used for verifying the application of GMM for its possibly first-time application on field measurement data. The identified VIV load dominated cases were used as criteria to evaluate the cluster results. It is the intention to reduce the need of manual evaluation. Even so, GMM still provides more details and better interpretability on the riser response due to different load combinations.

\newpage
\section{Cluster analysis of the Helland-Hansen data}
\subsection{Overview of input parameters}
\begin{figure}[!h]
 		\centering
         \includegraphics[width=1.\textwidth]{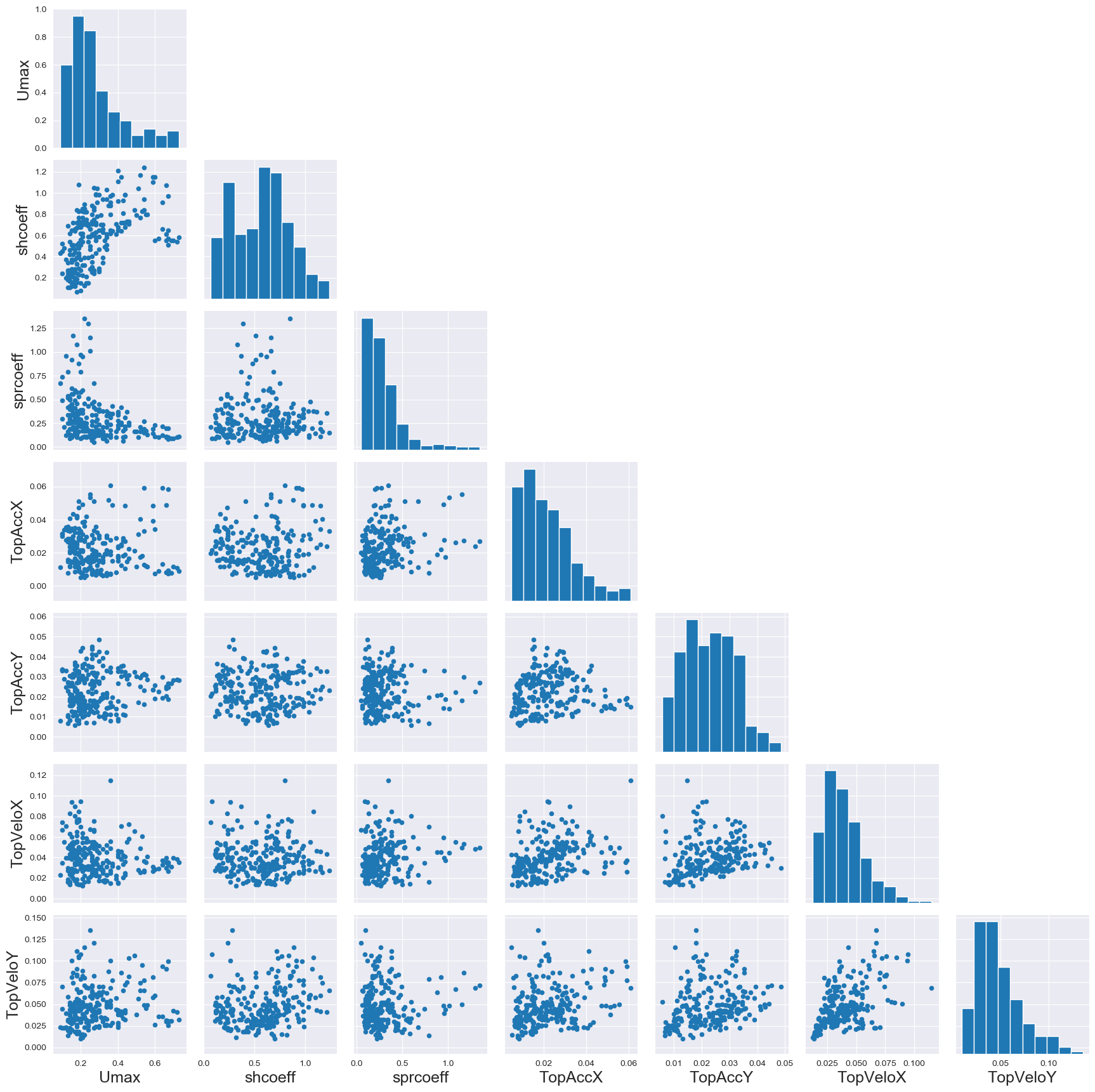}
         \caption{Distribution of parameters used in cluster analysis.}
         \label{fig:hist_para}
\end{figure}

A total of 7 parameters were selected representing environmental loads due to current, waves and vessel motions. They represent key load processes that can affect riser response. The distributions of cluster analysis input parameters are presented in Fig.\ \ref{fig:hist_para}. It is difficult to visualize this high dimensional data, and hence the data is presented by the histogram of each parameter and scatter plots between all pairs of parameters. The plot also reveals the relationships of some parameters, e.g., high current speed tends to be associated with high shearedness of the current profile. Larger directionality of the current was observed for low current speed. The number of parameters can be further increased, which may reveal more features, but it also makes the results interpretation more difficult. Significant overlap of these parameters can also be seen and visual separation of data groups is not feasible. The strong overlap of the data indicates that hard separation may be difficult. Gaussian mixture models are probabilistic models and it uses the soft clustering approach for distributing the points in different clusters.  

\begin{figure}[!h]
 		\centering
         \includegraphics[width=.98\textwidth]{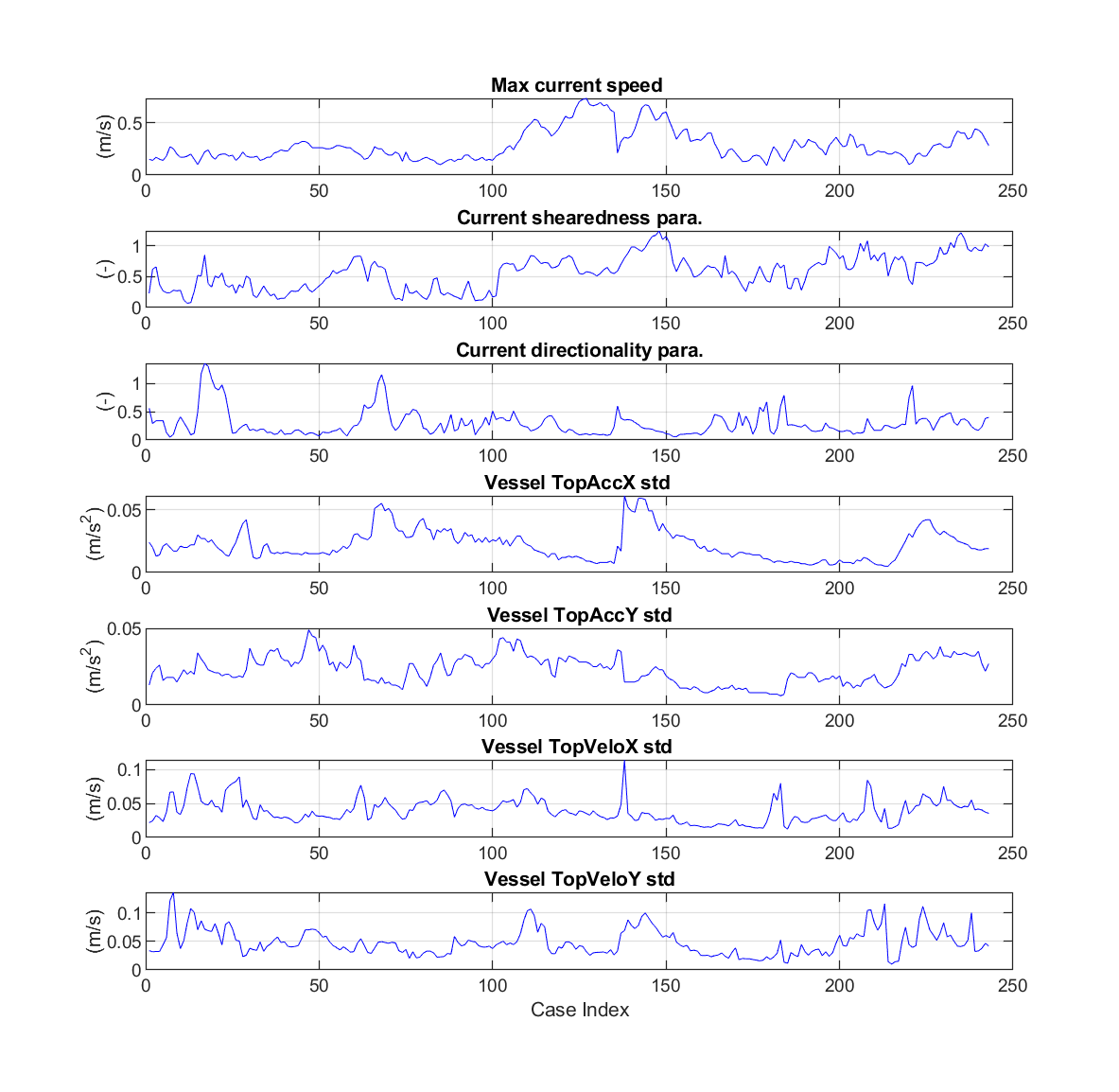}
         \caption{Variation of the environmental load parameters in chronological order.}
         \label{fig:Param_time}
\end{figure}

Some of the parameters are also presented in Fig.\ \ref{fig:Param_time}, as a function of the measurement in chronological order. Both gradual and sudden changes of weather conditions are observed and there are not two cases having identical conditions. 

\clearpage
\newpage
\subsection{Cluster analysis results}
\begin{figure}[!h]
 		\centering
         \includegraphics[width=.9\textwidth]{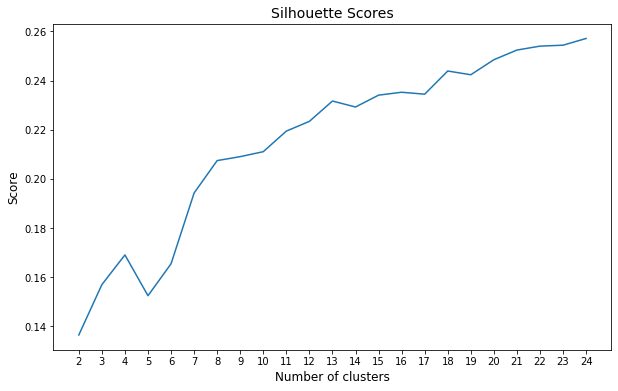}
         \caption{Silhouette score for various numbers of clusters.}
         \label{fig:Silhouette_12clusters}
\end{figure}

Silhouette score uses compactness of individual clusters (intra cluster distance) and separation amongst clusters (inter cluster distance) to measure an overall representative score of how well our clustering algorithm has performed. The silhouette score was calculated for different numbers of clusters, as presented in Fig.\ \ref{fig:Silhouette_12clusters}. It is seen that the silhouette score increases with the number of clusters, and we may conclude that a minimum of 8 clusters are needed to get a good score. The selection of the number of clusters is based on the compromise that the data is sufficiently separated with a minimum number of clusters. 

\begin{figure}[!h]
 		\centering
     \begin{subfigure}[b]{.7\textwidth}
         \centering
         \includegraphics[width=0.8\textwidth]{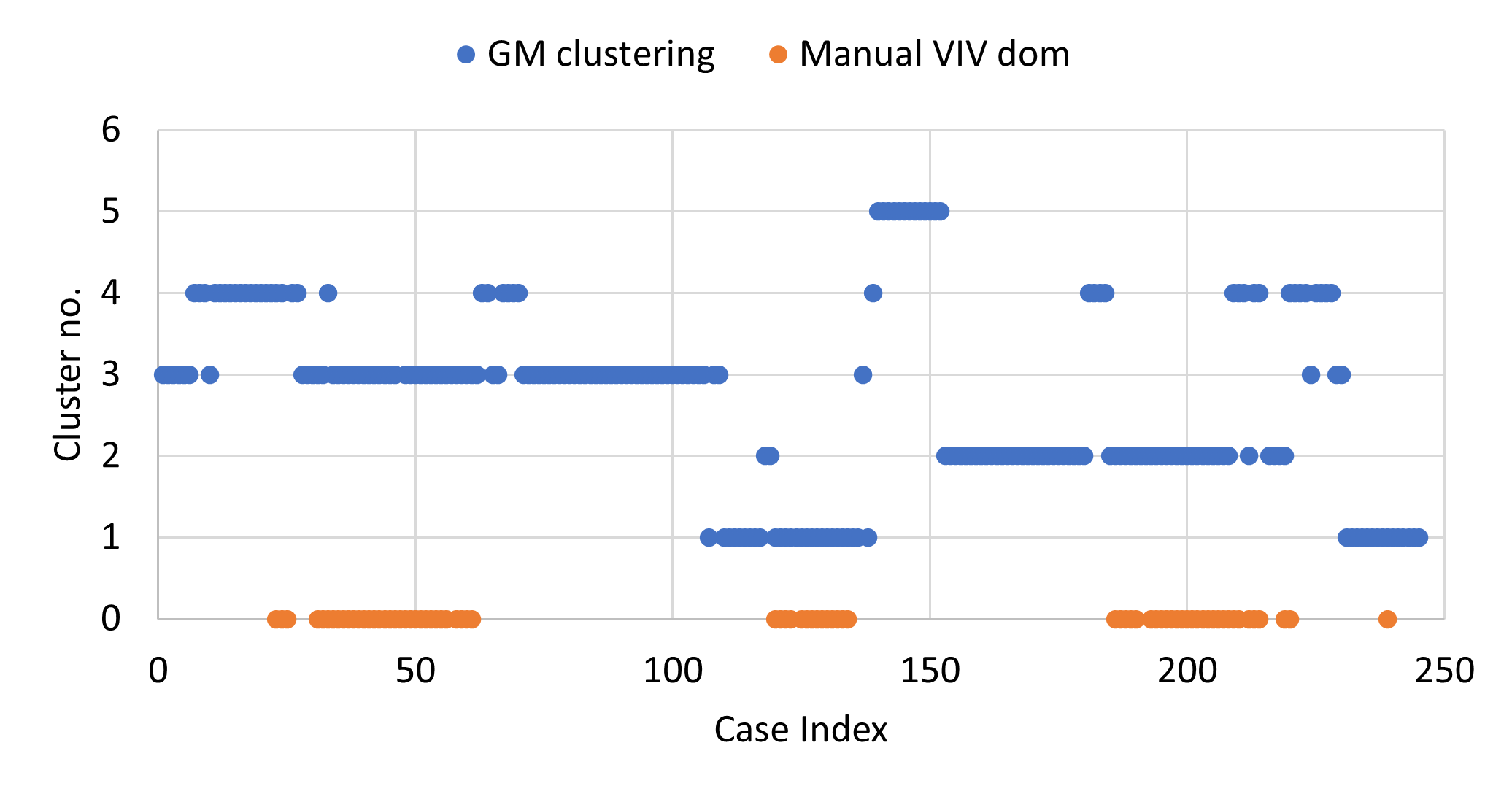}
         \caption{5 clusters}
     \end{subfigure}
     \hfill
     \begin{subfigure}[b]{.7\textwidth}
         \centering
         \includegraphics[width=0.8\textwidth]{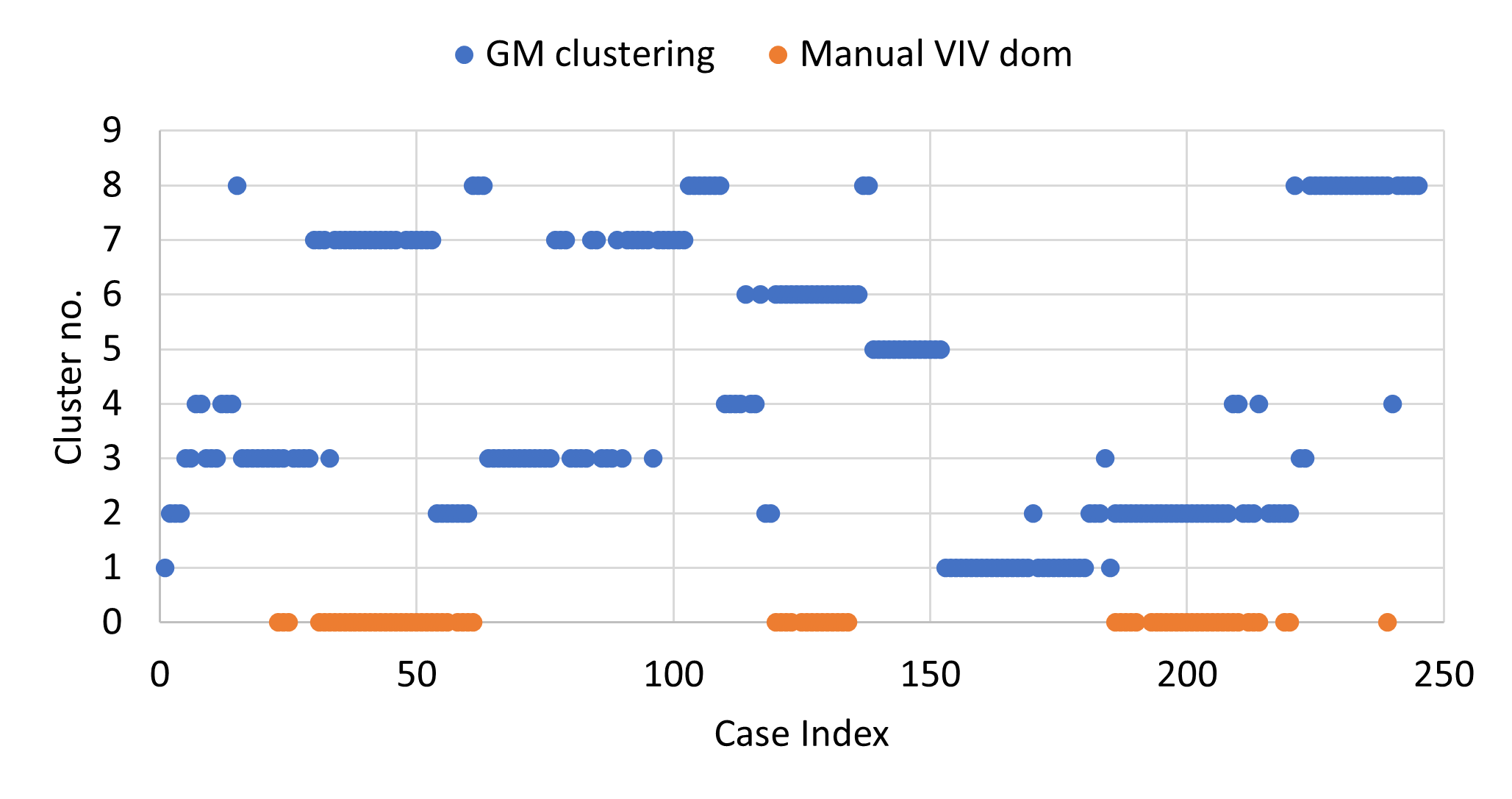}
         \caption{8 clusters }
     \end{subfigure}
     \hfill
     \begin{subfigure}[b]{.7\textwidth}
         \centering
         \includegraphics[width=0.8\textwidth]{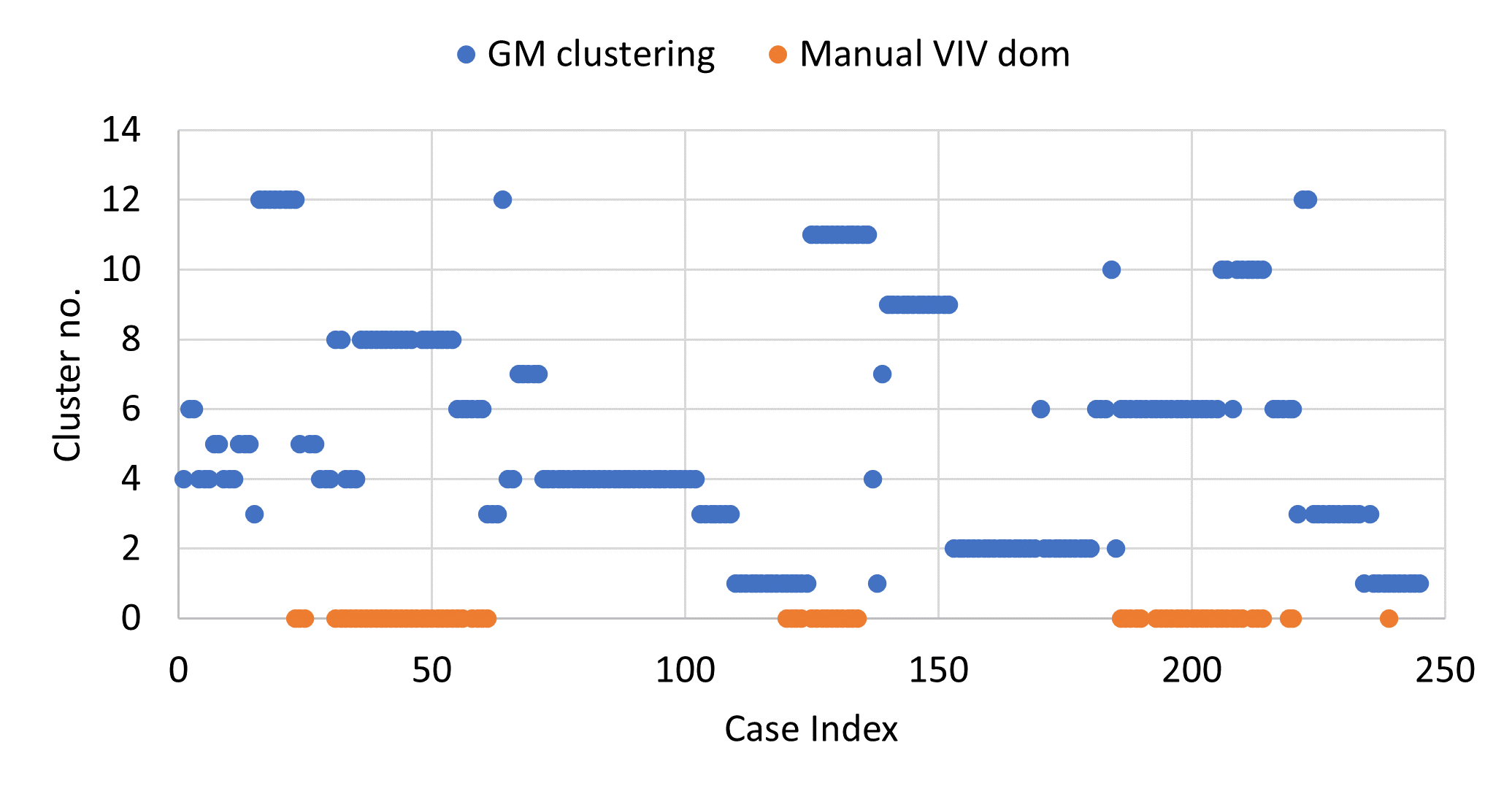}
         \caption{12 clusters}
     \end{subfigure}
     \hfill
     \begin{subfigure}[b]{.70\textwidth}
         \centering
         \includegraphics[width=0.75\textwidth]{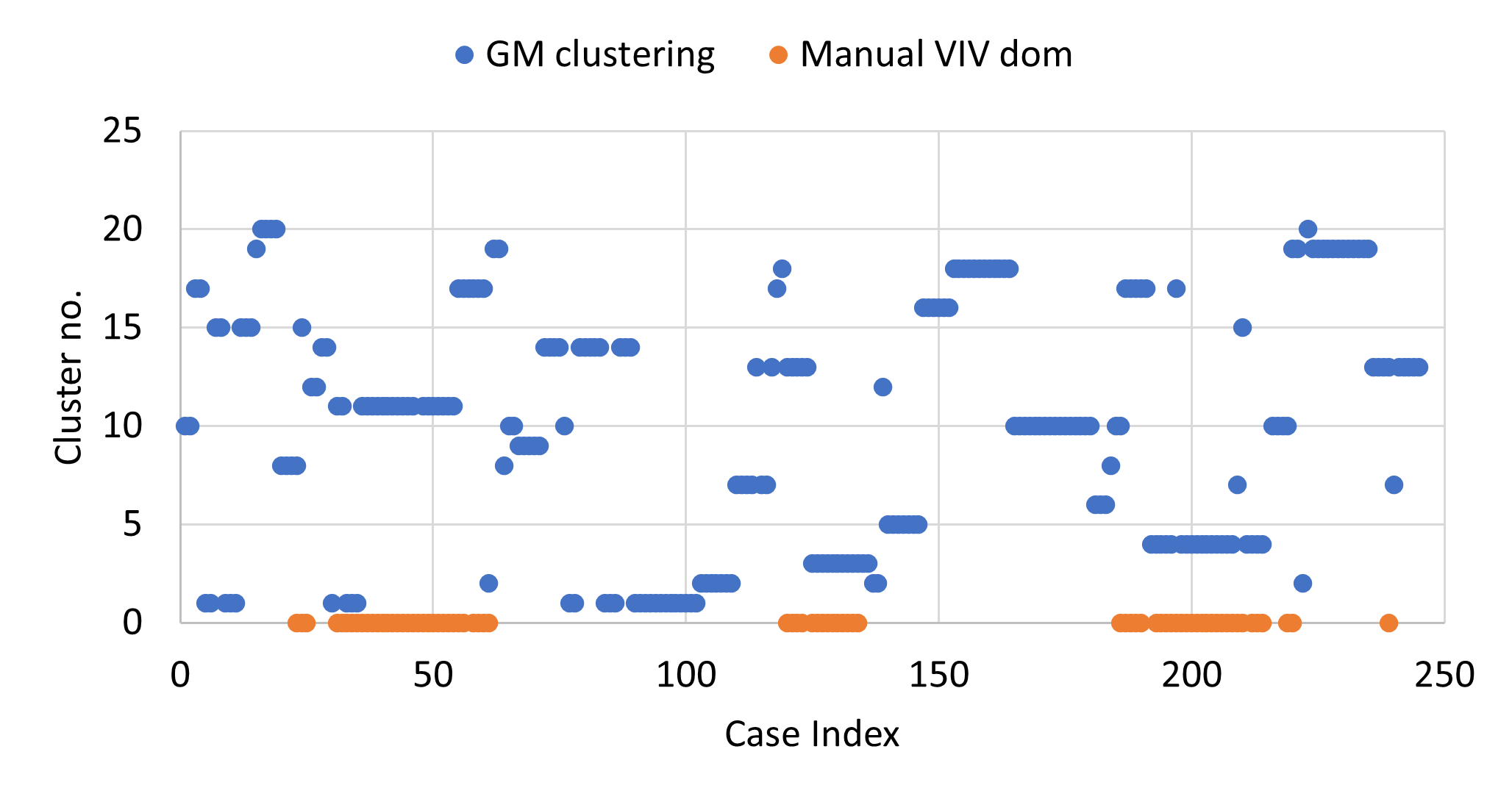}
         \caption{20 clusters}
     \end{subfigure}
     \hfill
\caption{Sensitivity on the selection of number of clusters}\label{fig:Sens_clusters}
\end{figure}

A sensitivity study on the selection of the number of clusters is presented in Fig.\ \ref{fig:Sens_clusters}. It can be seen that more differences can be identified with increasing number of clusters. However, the smaller clusters can be less representative due to the size of the data. The cases have also been evaluated manually based on domain knowledge. The frequency contents of the acceleration signals at sensors no.\ 3 and 4 were used to evaluate the characteristics of the riser responses. This may not reflect all aspects of riser responses, but it provides information based on local response measurements. The cases are shown to be dominated by VIV load (i.e., narrow-banded riser response dominated by the frequency close to Strouhal frequency). In addition, the identified VIV load dominated cases have different frequencies as they include cases with different current speed ranges.

\begin{figure}[!h]
 		\centering
         \includegraphics[width=.5\textwidth]{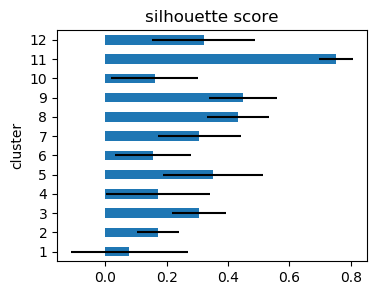}
         \caption{Silhouette score for individual clusters when 12 clusters are selected.}
         \label{fig:Silhouette_12_individual_cluster}
\end{figure}

\newpage
The manual cluster results were compared with the GM cluster results. 12 clusters were selected, at which point the gradient of the silhouette score trend becomes smaller. This means that less improvement is expected with further increase of the number of clusters. It can be seen that the manually identified VIV load dominating cases center around case no. 45, 125 and 200. This corresponds to GM cluster no. 6, 8, 10, 11 in Fig.\ \ref{fig:Sens_clusters}c. Some discrepancy can be seen: e.g., five cases in cluster no. 1 overlap with manual results. However, the cluster results using totally different methods and input parameters seem to be consistent. It is also noted that the silhouette score is an average of scores for individual clusters, as shown in Fig.\ \ref{fig:Silhouette_12_individual_cluster}. It can be seen that cluster no.\ 11 is best separated with the highest silhouette score and cluster no.\ 1 has the largest uncertainty, i.e. it is the cluster most likely be mixed with other clusters. The horizontal bar is a measure of the variation and indicates the sensitivity on the silhouette score due to random initialization. 

\begin{figure}[!h]
 		\centering
         \includegraphics[width=1.0\textwidth]{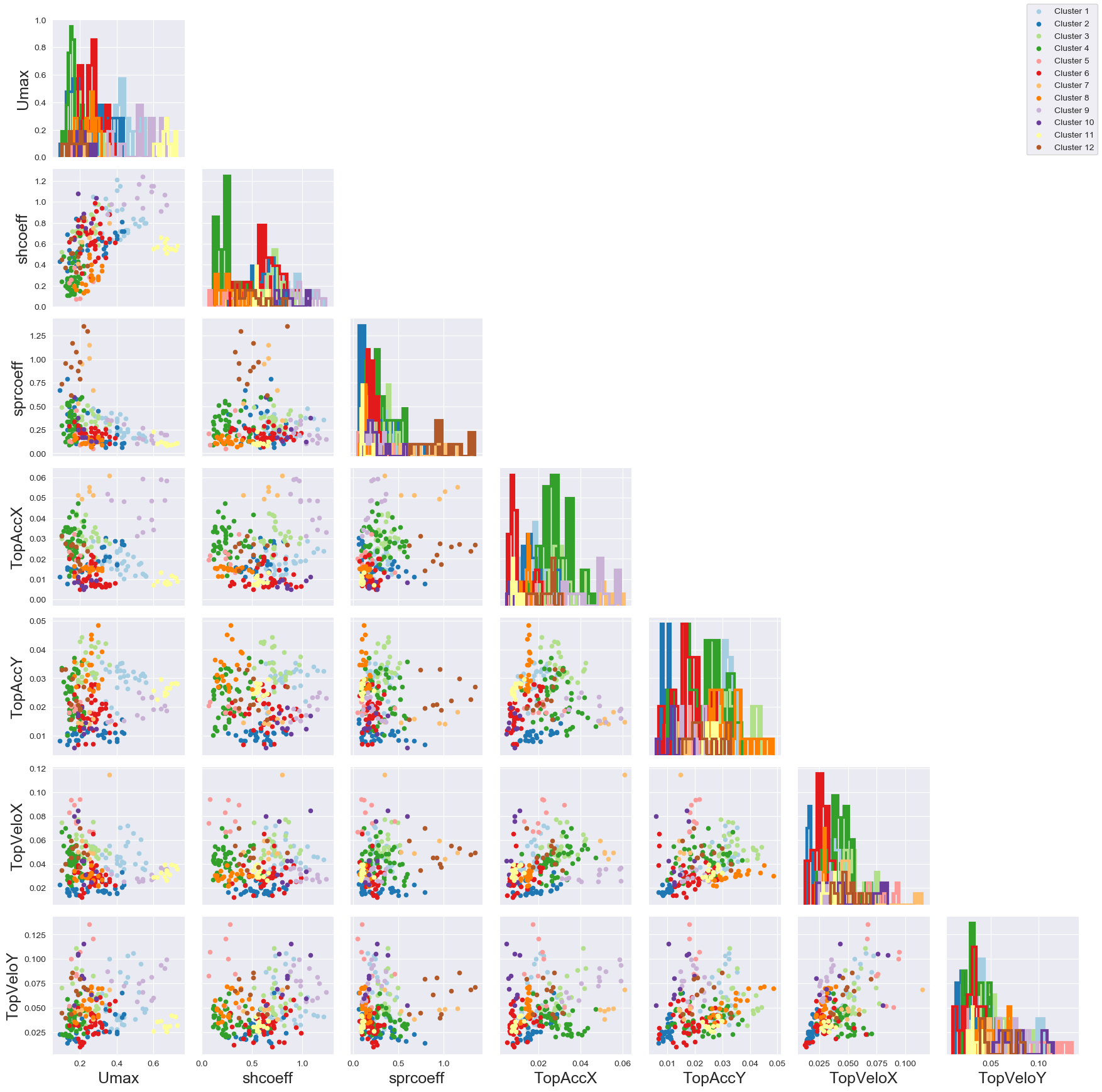}
         \caption{Distribution of 12 clusters}
         \label{fig:hist_12_clusters}
\end{figure}

The distribution of the 12 clusters is presented in Fig.\ \ref{fig:hist_12_clusters}. It can be seen that some of the identified clusters seem to be separated from the others based on certain parameters. For example, cluster no.\ 12 has a high spreading coefficient and low current speeds. In general, however, the clusters have strong overlaps in this high dimensional parameter space. 

\clearpage
\newpage

\begin{figure}[!h]
 		\centering
     \begin{subfigure}[b]{.45\textwidth}
         \centering
         \includegraphics[width=0.7\textwidth]{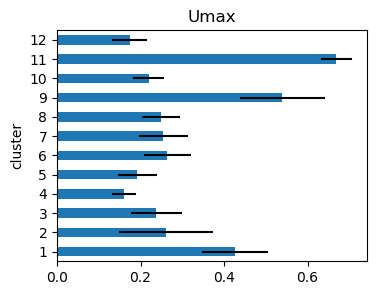}
         \caption{Maximum current speed  }
         \label{fig:12_Umax}
     \end{subfigure}
     \hfill
     \begin{subfigure}[b]{0.45\textwidth}
         \centering
         \includegraphics[width=0.7\textwidth]{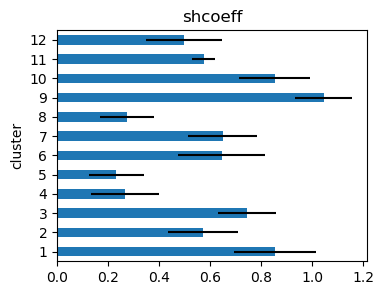}
         \caption{Current shearedness }
         \label{fig:12_shcoeff}
     \end{subfigure}
     \hfill
     \begin{subfigure}[b]{0.45\textwidth}
         \centering
         \includegraphics[width=0.7\textwidth]{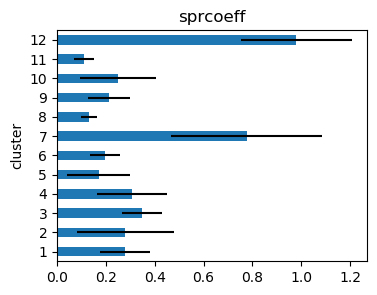}
         \caption{Current directionality  }
         \label{fig:12_sprcoeff}
     \end{subfigure}
     \hfill
     \begin{subfigure}[b]{0.45\textwidth}
         \centering
         \includegraphics[width=0.7\textwidth]{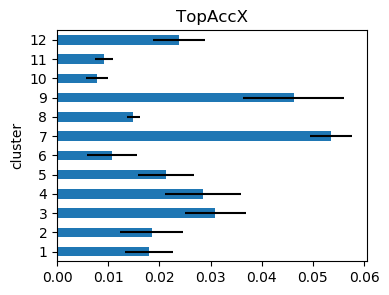}
         \caption{Vessel acceleration in the main current direction }
         \label{fig:12_TopAccX}
     \end{subfigure}
     \hfill
     \begin{subfigure}[b]{0.45\textwidth}
         \centering
         \includegraphics[width=0.7\textwidth]{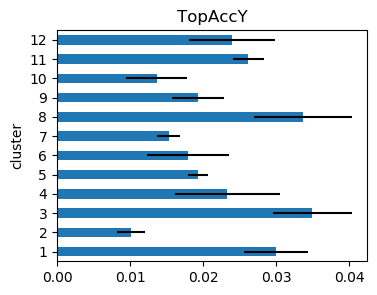}
         \caption{Vessel acceleration normal to the main current direction}
         \label{fig:12_TopAccY}
     \end{subfigure}
     \hfill
     \begin{subfigure}[b]{0.45\textwidth}
         \centering
         \includegraphics[width=0.7\textwidth]{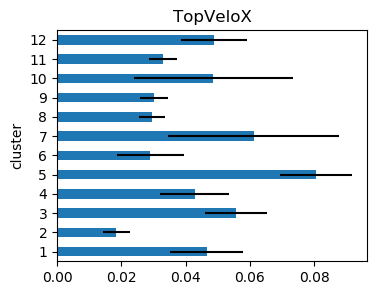}
         \caption{Vessel velocity in the main current direction }
         \label{fig:12_TopVeloX}
     \end{subfigure}
     \hfill
     \begin{subfigure}[b]{0.45\textwidth}
         \centering
         \includegraphics[width=0.7\textwidth]{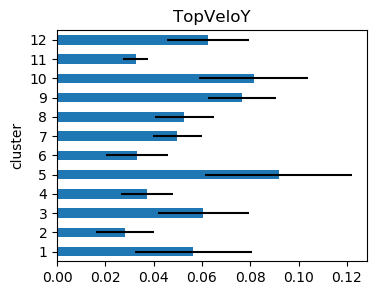}
         \caption{Vessel velocity normal to the main current direction}
         \label{fig:12_TopVeloY}
     \end{subfigure}
     \hfill
\caption{Mean and standard deviation of parameters in each cluster}\label{fig:Para_distr_12_clusters}
\end{figure}

The characteristics (i.e., mean and standard deviation) of the environmental parameters in individual clusters are presented in Fig.\ \ref{fig:Para_distr_12_clusters}. 

\newpage
The mean values are also presented in Table \ref{tab:Riser_response cluster} and marked with colors indicating the levels of magnitude, i.e., red for the highest value, brown for medium value and green for the lowest value. We observe that there are 12 cases in cluster no.\ 11. The mean value of the maximum current speed of the 12 cases is about 0.67 m/s, which is highest compared to that of other clusters, which corresponds to a Strouhal frequency about 0.15 Hz. This is in the same range as the wave frequency and corresponding to the fifth mode. The actual response frequency will be affected by other parameters as well. The averaged shearedness is 0.57 and in the medium level, while the directionality of the current, the wave loads (vessel acceleration), and vessel motion induced velocity are relatively low. Cluster no.\ 4 has the lowest averaged current speed; the wave loads and vessel induced velocity are relatively low as well. The current profile in cluster no.\ 12 has the highest directionality and low current speed. Cases in cluster no.\ 2, 6 and 8 have medium current speeds and low wave loads and vessel motion induced velocity.  In summary, environmental load conditions are separated in 12 clusters and the riser responses in each of the clusters are expected to be similar. 

\begin{table}[!h]
    \centering
    \caption{Riser response characteristics of each cluster}
    \includegraphics[width=1\textwidth]{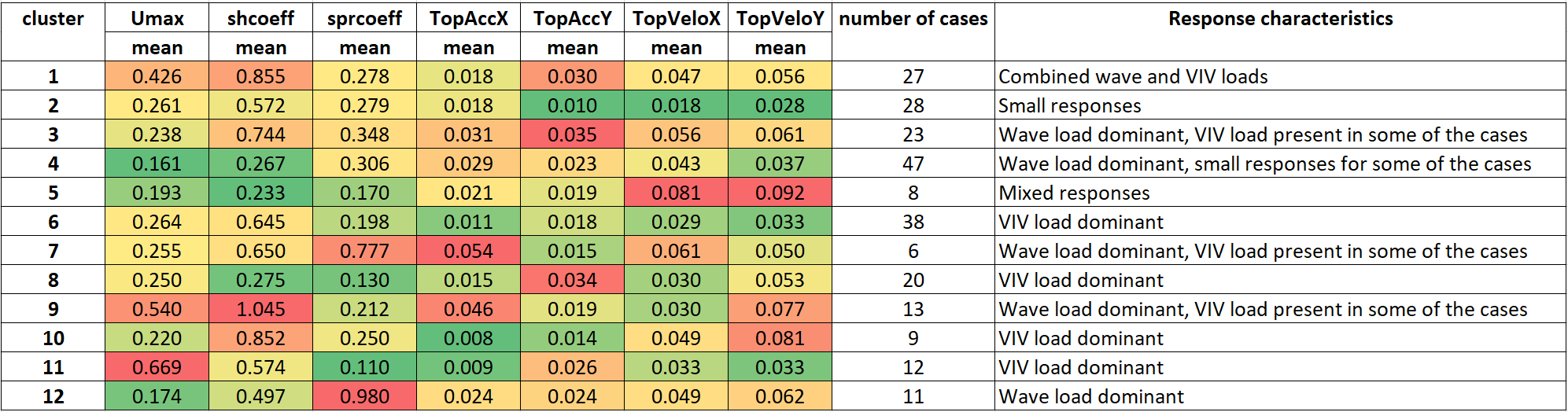}
\label{tab:Riser_response cluster}
\end{table}

The characteristics of the riser response of each cluster are also summarized in Table \ref{tab:Riser_response cluster}. In general, clusters are able to separate different riser response characteristics. The differences in riser responses are also correlated to the corresponding environmental load parameters. 

\begin{figure}[!h]
 		\centering
         \includegraphics[width=.98\textwidth]{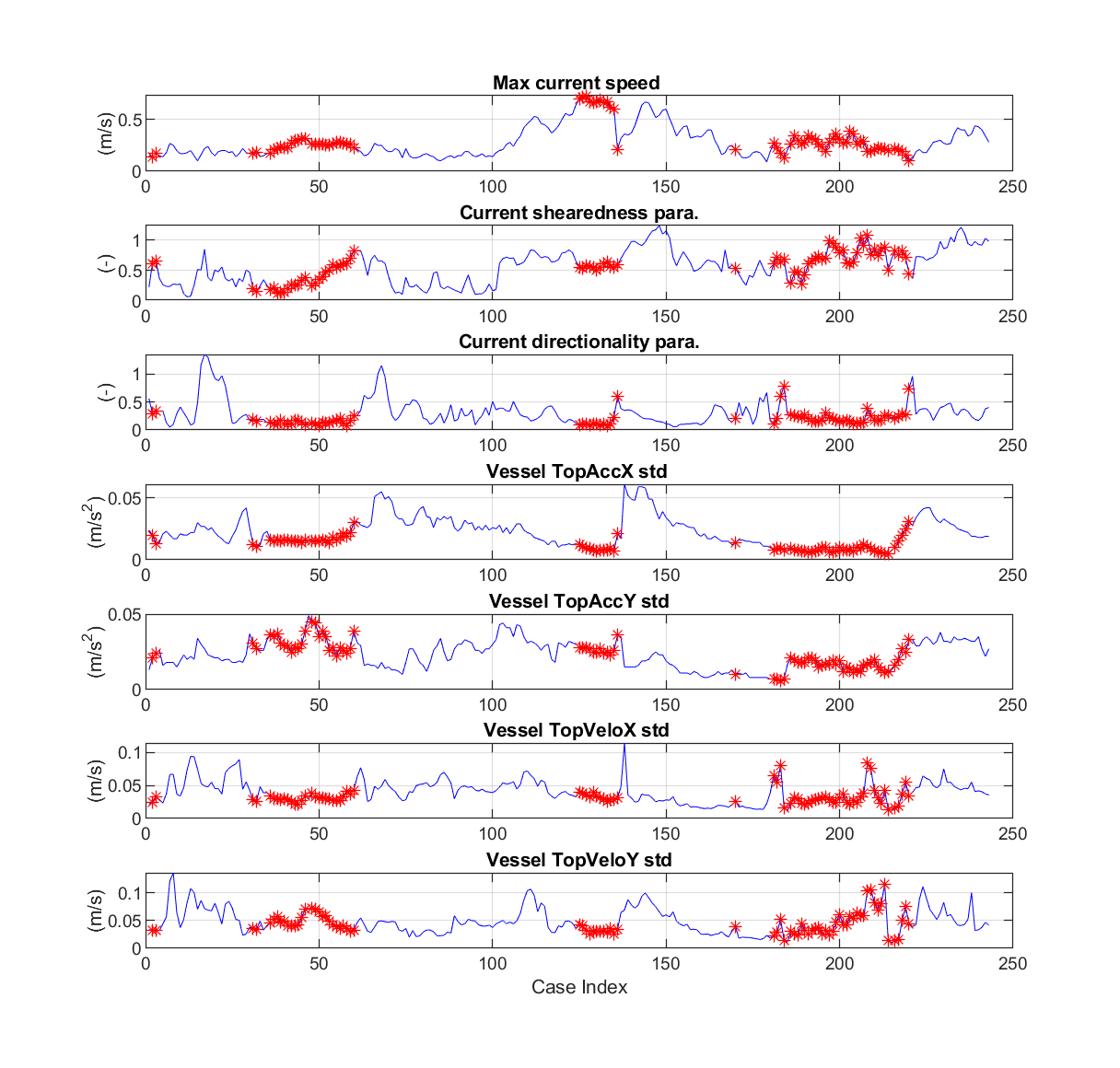}
         \caption{Environmental load parameters corresponding to VIV load dominating cases}
         \label{fig:Param_time_VIV}
\end{figure}

Cases in cluster no.\ 6, 8, 10 and 11 are in general dominated by VIV load, which are related to 1) high current speed and medium/low wave loads (cluster no.\ 11); 2) medium current speeds and relatively low wave loads (cluster no.\ 6 and 8); 3) low current speed, but high vessel motion induced velocity (cluster no.\ 10). Wave loads normal to the main current direction in cluster no.\ 8 do not seem to disturb VIV responses. Otherwise, VIV frequency are generally not present for low current speed cases as indicated in clusters no.\ 4 and 12. In such cases, the responses are small or dominated by the wave frequency. The associated environmental load parameters for the cases dominated by VIV loads are presented in Fig.\ \ref{fig:Param_time_VIV}. It is seen that these cases are in general associated with medium and high current speeds, low current directionality and small wave loads in the main current direction. In some cases, vessel motion induced velocity is of the same order of the current speed, e.g., case 213.

\begin{figure}[!h]
 		\centering
         \includegraphics[width=.4\textwidth]{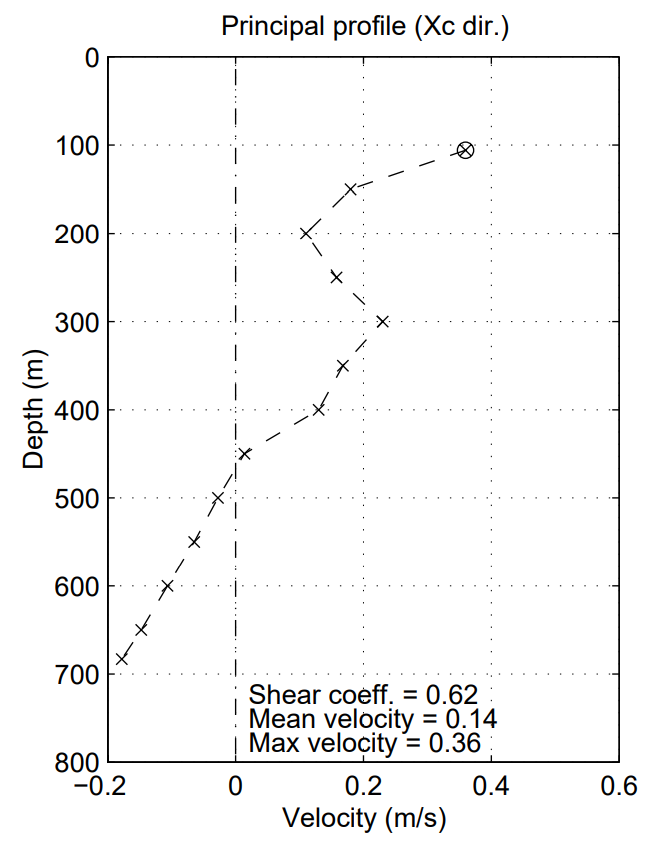}
         \caption{Current profile for case 159 in cluster no.2}
         \label{fig:Uc_052802_159}
\end{figure}

VIV load and wave load will compete or interact for cases in clusters no.\ 1, 3, 7 and 9, when the current speed is about the medium level.
Cases in cluster no.\ 2 have medium current, mild shearedness and directionality as well as relatively low wave loads, which seems to be ideal for the presence of VIV responses as in cluster no.\ 6. However, only small responses are observed. The further investigation revealed that the current profile of these cases are special and cannot be properly represented by the shearedness parameter. One example is shown in Fig.\ \ref{fig:Uc_052802_159}. The maximum speed is 0.36 $\mathrm{m}/\mathrm{s}$, but the current direction turns $180^{\circ}$ at a water depth of 450 m. The shearedness parameter was calculated with the absolute value of the current speed, which means that the actual local current speed variation is much higher than the calculated value. It is known that quick changes in the local current speeds make it difficult for vortex shedding to be synchronized along the length of the riser, which can be the reason for small responses.

\clearpage
\newpage
\section{Time domain {VIV} prediction model}
The riser responses of the selected 242 full scale measurement events are predicted using the time domain VIV responses analysis tool: VIVANA-TD. We introduce first the theory background of VIVANA-TD, then the predicted riser responses including maximum displacement standard deviation along the riser and the corresponding dominant response frequencies. The responses are presented with respect to the 12 identified clusters. The prediction accuracy for different clusters varies and the possible reasons are explored and discussed with consideration of the characteristics of individual cluster and the limitations of the current time domain VIV calculations.

\subsection{Hydrodynamic load formulation}
The hydrodynamic load model combines a vortex shedding force term with Morison's equation based on the strip theory. The cross-sectional force of the cylinder are illustrated in Fig.\ \ref{fig:TD_VIV_Vector}, and the hydrodynamic forces on the cylinder (per unit length) are represented by Eq. (\ref{Eq.TDVIV}) \citep {KIM2022103057}:

\begin{figure}[!h]
 		\centering
         \includegraphics[width=.7\textwidth]{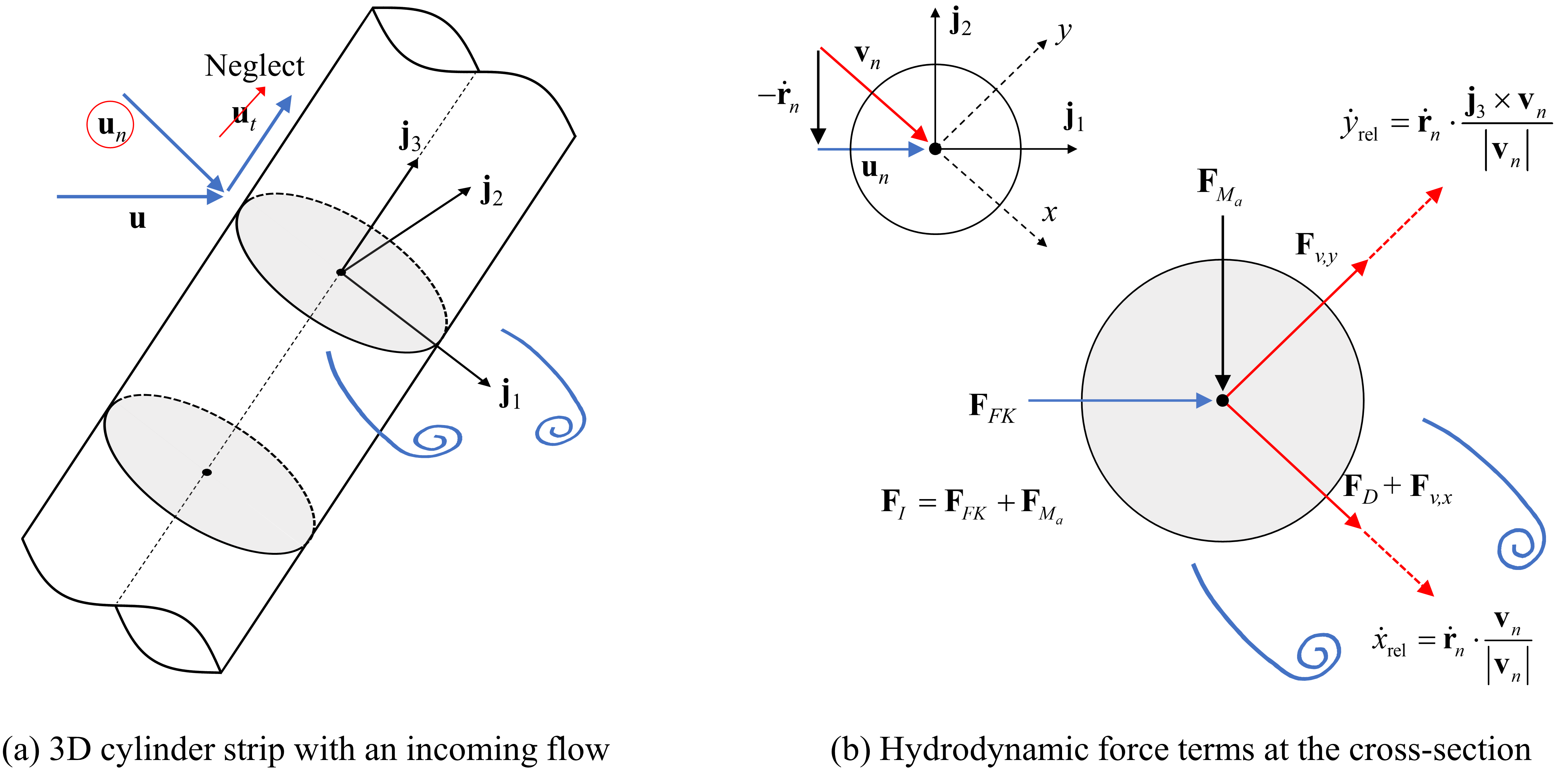}
         \caption{VIV force coordinate system \citep{KIM2022103057}}
         \label{fig:TD_VIV_Vector} 
\end{figure}

\begin{equation}
\begin{array}{l}
{{\bf{F}}_n} = \,\underbrace {\underbrace {{C_M}\rho \frac{{\pi {{D}^2}}}{4}{{{\bf{\dot u}}}_n}}_{{\mathop{\rm Froud - Kriloff\,\,force}}\,\,({{\bf{F}}_{FK}})} - \underbrace {({C_M} - 1)\rho \frac{{\pi {{D}^2}}}{4}{{{\bf{\ddot r}}}_n}}_{{\mathop{\rm Added\,\,mass\,\,force}}\,\,({{\bf{F}}_{{M_a}}})}}_{{\mathop{\rm Inertia}\nolimits} \,{\mathop{\rm force}\nolimits} \,\,({{\bf{F}}_I})} + \underbrace {\frac{1}{2}\rho {D}{C_D}\left| {{{\bf{v}}_n}} \right|{{\bf{v}}_n}}_{{\mathop{\rm Drag\,\,force}}\,\,({{\bf{F}}_D})}\\
\\
\,\,\,\,\,\, + \underbrace{\underbrace{\frac{1}{2}\rho {D}{C_{v,x}}\left| {{{\bf{v}}_n}} \right|{{\bf{v}}_n}\cos {\phi _{{\rm{exc}},x}}}_{{\mathop{\rm IL\,\,vortex\,\,shedding\,\,force}}\,\,({{\bf{F}}_{v,x}})} + \underbrace {\frac{1}{2}\rho {D}{C_{v,y}}\left| {{{\bf{v}}_n}} \right|({{\bf{j}}_3} \times {{\bf{v}}_n})\cos {\phi _{{\rm{exc}},y}}}_{{\mathop{\rm CF\,\,vortex\,\,shedding\,\,force}}\,\,({{\bf{F}}_{v,y}})}}_{{\mathop{\rm Vortex\,\,shedding\,\,force}}}
\end{array}
\label{Eq.TDVIV}
\end{equation}
where, $C_M$, $C_D$, $D$, and $\rho$ are the inertia coefficient, drag coefficient, cylinder diameter, and fluid density, respectively. $C_{v,x}$ and $C_{v,y}$ are in-line and cross-flow vortex shedding force coefficients; ${\phi_{{\rm exc},x}}$ and ${\phi_{{\rm exc},y}}$ are the in-line and cross-flow instantaneous phases of the vortex shedding forces. ${{\bf u}}_n$  is normal to the cylinder axis as shown in Fig.\ \ref{fig:TD_VIV_Vector}. Each force term of the load model is defined in the direction of the incoming flow velocity, ${{\bf u}}_n$, the structure local response, ${{\bf r}}_n$, and the relative flow velocity, ${{\bf v}}_n$ (= ${{\bf u}}_n-{{\bf{\dot r}}_n}$). 

The first three force terms are the inertia force $F_I$ ($=F_{FK}+F_{M_{a}}$) and drag force $F_D$, while $F_{v,x}$ and $F_{v,y}$ are in-line and cross-flow vortex shedding forces. The cross-flow vortex shedding force represents a lift force in the local $y$ direction, which is the direction normal to the relative flow vector $v_n$. The in-line vortex shedding force $F_{v,x}$ describes a fluctuating drag force in a local $x$ direction, which is in the same direction as $v_n$. 

\subsection{Synchronization of vortex shedding forces}
The synchronization model describes the phase-coupling between force and response to obtain lock-in. The instantaneous frequency of the vortex shedding force will be adjusted so that the phase of the force can maintain lock-in with the phase of the cylinder velocity. Part of the vortex shedding force will be in phase with velocity, which determines the energy input to the structure. The hydrodynamic damping is determined by the drag load term. The rest of the vortex shedding force is in phase with acceleration and contributes to the added mass term. CF synchronization model is as follow:

\begin{equation}
\frac{{d{\phi _{{\mathop{\rm exc}\nolimits} ,\,y}}}}{{dt}} = 2\pi {f_{{\mathop{\rm exc}\nolimits} ,y}} = \frac{{2\pi \left| {{{\bf{v}}_n}} \right|}}{D}{\widehat f_{{\mathop{\rm exc}\nolimits} ,\,y}}
\label{Eq.6}
\end{equation}
\begin{equation}
{\widehat f_{{\mathop{\rm exc}\nolimits} ,\,y}} = {\widehat f_{0,\,y}} + \Delta {\widehat f_y}\sin{\theta_{y}}
\label{Eq.7}
\end{equation}
where, $\Delta f_{\mathop{y}}$, and $\hat f_{0,{y}}$ determine the cross-flow synchronization range. ${\theta_{y}}$ is the phase difference between the cylinder cross-flow velocity, $\phi _{{\dot y}_{{\mathop{\rm rel}\nolimits}}}$ and the cross-flow vortex shedding force, $\phi _{{\mathop{\rm exc}\nolimits},{\rm y}}$, ($\theta_{y}= {{\phi _{{{\dot y}_{{\mathop{\rm rel}\nolimits} }}}} - {\phi _{{\mathop{\rm exc}\nolimits},{y}}}}$). ${{\phi _{{{\dot y}_{{\mathop{\rm rel}\nolimits} }}}}}$ can be numerically approximated at each time step \citep {MT2017}. $\hat f_{0,{y}}$ refers to the point of full synchronization, where the vortex shedding force is almost completely in phase with cylinder velocity or ${\theta_{y}=0}$. 

The IL synchronization model \citep{Kim2021OE} has a similar formulation as CF model. IL response frequency is assumed to be twice that of CF frequency, which is normally the case for marine risers.

\subsection{Establish empirical parameters to be used in VIVANA-TD}
It is well known that hydrodynamic coefficients are strongly affected by Reynolds (Re) effects. The Re number of the Helland-Hansen riser ranges from 1 to 7e5 based on the maximum current speed and the diameter of the buoyancy element. 

A review of the available test data \citep{Ding2004, Potts2018, HighReAD2020, lie2013, yin2018b} showed variation of responses is dependent on the Reynolds number, roughness and model test set-up, i.e. mass ratio, structural damping, etc. The highest displacement amplitude ratio is in the range of 0.5D - 0.8D depending on roughness for $\mathrm{Re} >1e5$. Higher displacement up to 1.8D of a smooth cylinder was also observed close to $\mathrm{Re}=1e5$, which was considered not relevant for marine riser application. Strouhal number up to 0.25 - 0.3 is observed with a large scatter of data. The empirical parameter set defined in Table \ref{Table:Empirical_parameters} is proposed based on the evaluation of realistic VIV response amplitude and frequency at high Re.

\begin{table}[h]
\centering
\caption{Empirical hydrodynamic parameters} \label{Table:Empirical_parameters}
\begin{center}
\scalebox{0.9}{
\begin{tabular}{c c c c c c c c c } 
\hline
Parameters & $C_D$ & $C_M$ & $C_{v,y}$ & $C_{v,x}$ & {$\hat f_{0,y}$} & {${\hat f_y} range$} & {$\hat f_{0,x}$} & {$ {\hat f_x} range $}  \\
\hline
Value & 1.0 & 2.0 & 0.8 & 1.2 & 0.25 & 0.125 - 0.4 & 0.5 & 0.25 - 0.75\\
\hline
\end{tabular}}
\end{center}
\end{table}

\begin{figure}[!h]
    \centering
        \includegraphics[width=0.67\textwidth]{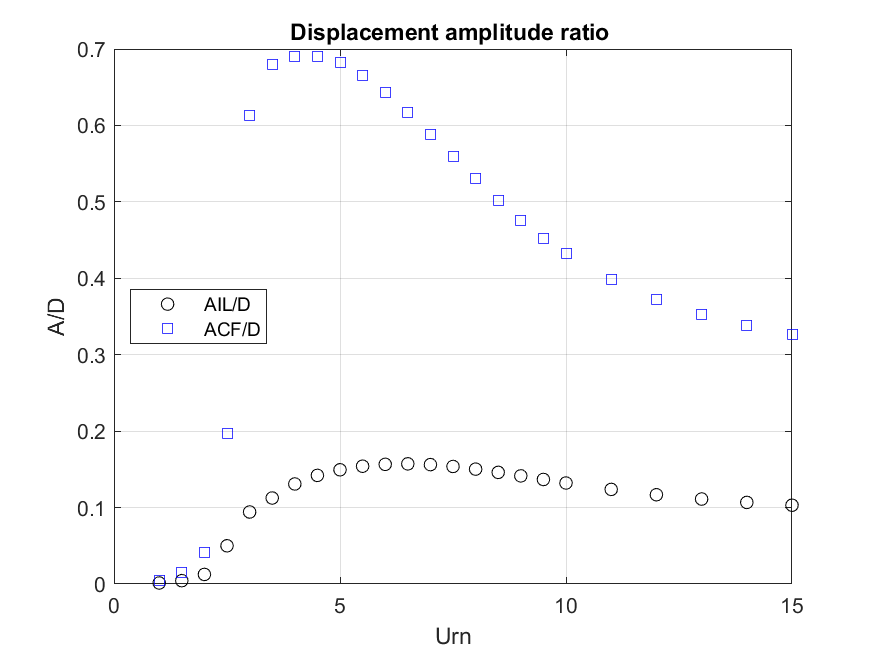}
        \caption{Predicted A/D based on a simulated elastic supported rigid cylinder}\label{fig:EP_AD_Urn}
\end{figure}

The effect of using this empirical parameter set was evaluated by simulation of a rigid cylinder. The simulated cylinder has a mass ratio of 2.0, ratio of the natural frequency in CF and IL direction is 2.0. The predicted displacement amplitude ratio ($A/D=\sqrt{2}A_{rms}/D$) is plotted against the reduced velocity ($U_{rn}=U/f_nD$) in Fig.\ \ref{fig:EP_AD_Urn}. The maximum cross-flow displacement amplitude ratio is 0.69, which corresponds to normalized frequency ($f_{hat}=f_{osc}D/U$) of 0.25. 

\newpage
\subsection{Riser response prediction}
\begin{figure}[!h]
 		\centering
         \includegraphics[width=1.\textwidth]{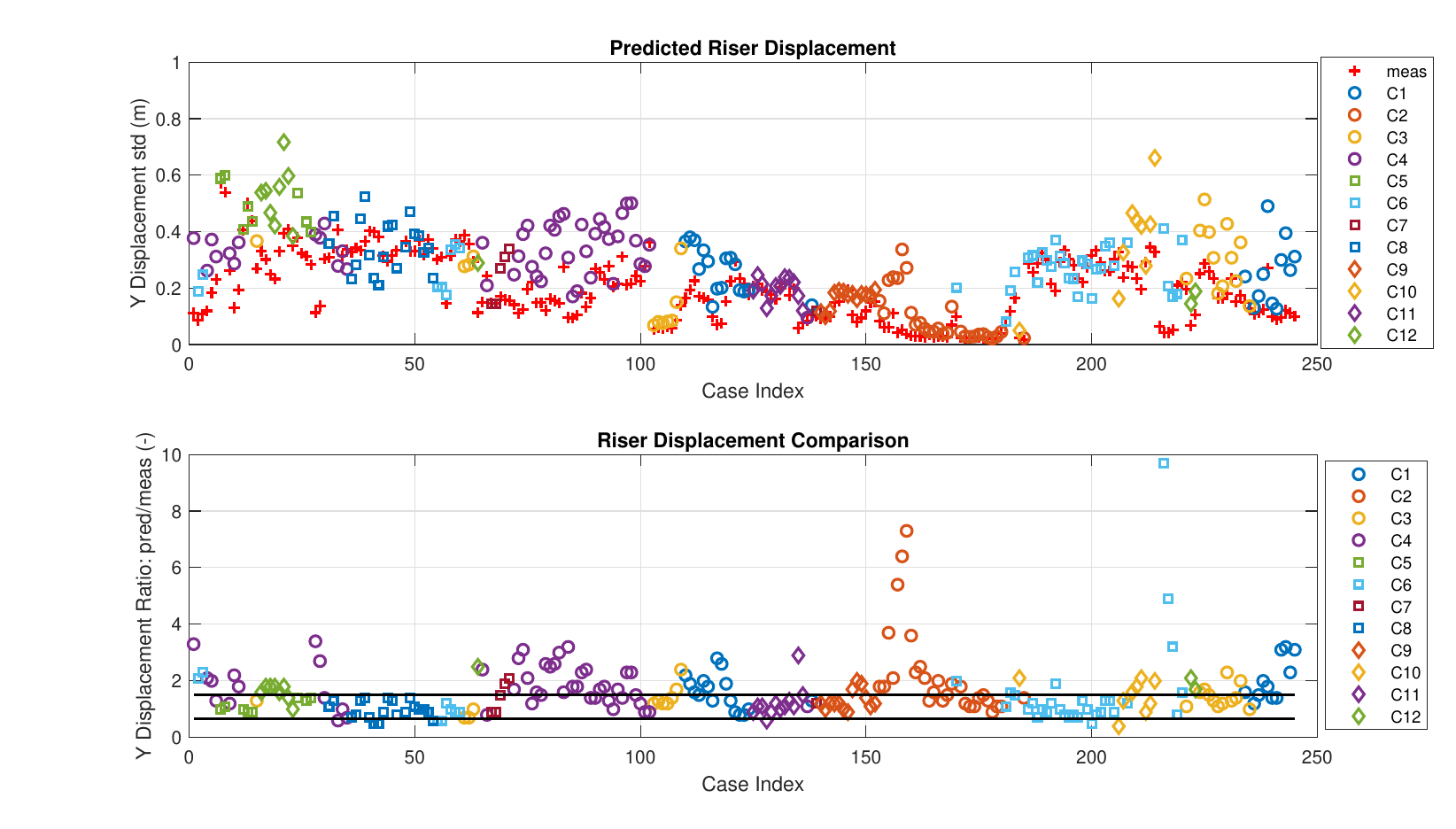}
         \caption{Comparison of displacement prediction. The measured displacement is compared with prediction organized in 12 clusters (C1-C12)}
         \label{fig:Disp_pred}
\end{figure}

\begin{figure}[!h]
 		\centering
         \includegraphics[width=1.\textwidth]{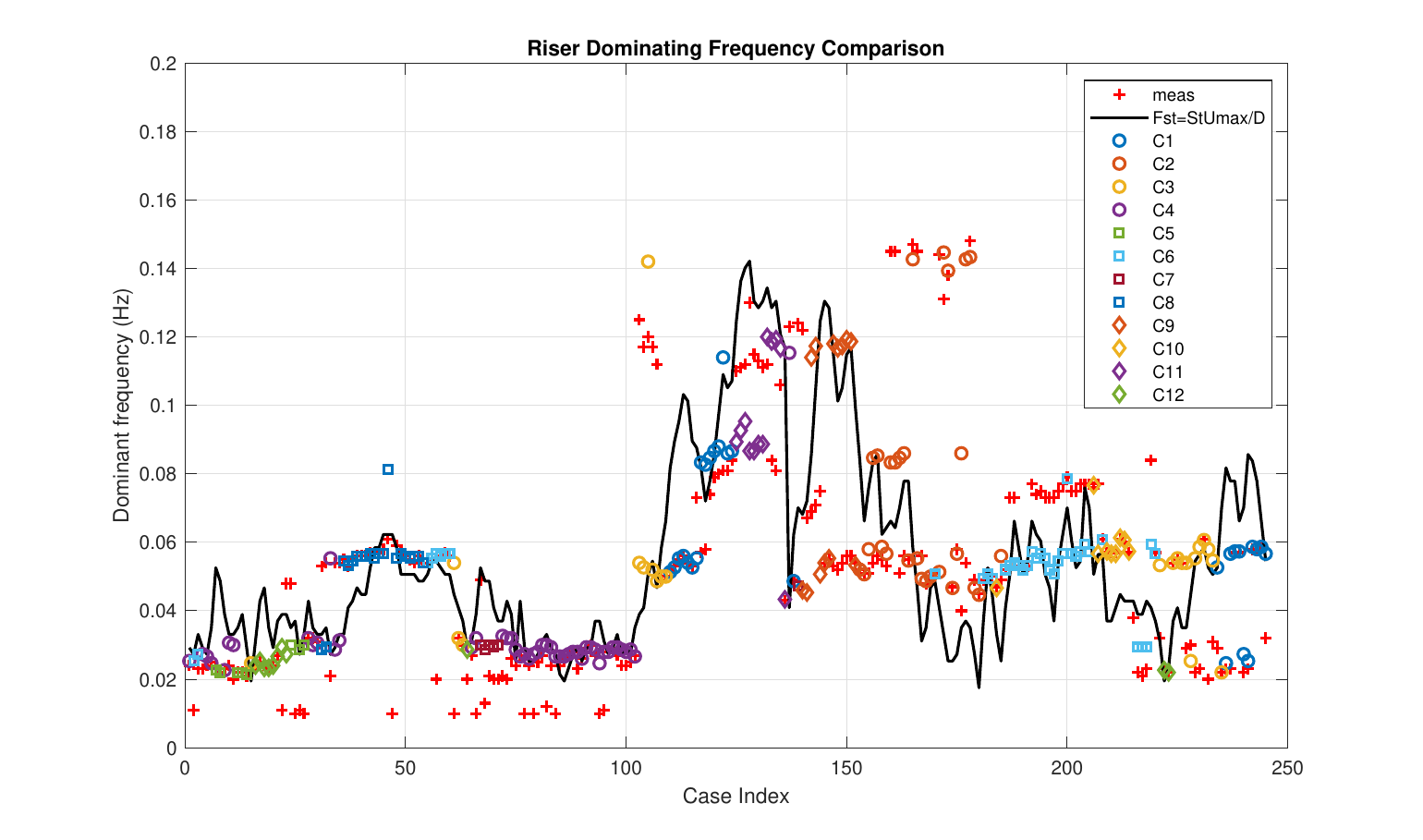}
         \caption{Comparison of dominant frequency.The measured frequency is compared with prediction organized in 12 clusters (C1-C12) and the estimated Strouhal frequency}
         \label{fig:Freq_pred}
\end{figure}

The actual measured 3-dimensional current profiles and estimated vessel motions were used in the simulation. The predicted maximum displacement standard deviation along the length of the riser and the corresponding dominating response frequency are compared with measurements in Fig.\ \ref{fig:Disp_pred} and Fig.\ \ref{fig:Freq_pred}. The results are presented in different clusters. The percentage of cases having a large error in the predicted maximum displacement (over-prediction or under-prediction larger than a factor of 1.5) and dominant frequency (difference larger than 0.03 Hz or miss-prediction of one mode) compared to measurement were also calculated and presented in Table \ref{table:Pred_eval}. In general, the prediction agrees well with the measurement. The deviation between predicted and measured displacement is within a factor of 1.5 for 141 among the total 242 cases (marked by the black lines in the figure). Under-prediction of displacement larger than a factor of 1.5 is seen for 10 cases. It is clear to see that VIV prediction accuracy varies for different environmental load combinations. Response prediction of VIV load dominating cases is good for clusters no.\ 6, 8, 10 and 11 in terms of displacement and/or frequency, which means that 70 \% of the cases have small displacement error and 94 \% of the cases have small frequency error. Larger deviations are observed for clusters no.\ 1, 4 and 12, which experience low current speeds and responses are dominated by wave loads. This deviation is expected as the wave loads were not included in the present simulation due to lack of monitored data. Such deviation seems to be smaller for higher current speed cases in clusters no.\ 3 and 9. Influence of wave loads on the prediction accuracy are reduced in such cases. Some of the cases in clusters no.\ 2 and 6 show large over-prediction; current changes direction in these cases, as seen in Fig.\ \ref{fig:Uc_052802_159}. The measured frequency is very low in some of the cases in cluster no.\ 4, e.g., between case no. 50 - 100. This is partially related to small and chaotic responses under small environmental loads. The estimated Strouhal frequency based on the maximum current speed gave an indication of the expected response frequency. The prediction discrepancy is related to 1) wave loads are not included in the simulation; 2) interaction of wave and VIV load processes requires different hydrodynamic parameters.    

Further improvement on the prediction can be achieved by including wave loads and applying adaptive hydrodynamic parameters \citep{wu2020improved} based on the different environmental load conditions indicated by the identified clusters. Optimization algorithm can be applied to obtain optimal hydrodynamic parameters based on the field measurements \citep{yin2022b}. The obtained empirical hydrodynamic parameters also need to be further consolidated based on extensive analysis of both field and laboratory test data. The internal flow of the drilling mud was not considered in this work, as it is not in present VIV prediction practice. However, several numerical studies \citep{THORSEN20191, DUAN2021103094}  show potential impact of internal flow on riser responses which may be considered in further analysis.

\clearpage
\begin{table}[!h]
\centering
\caption{Prediction evaluation}
\begin{tabular}{|c|c|c|c|}
\hline
Cluster no. & Number of cases & Large displacement prediction error (\%) & Large frequency prediction error (\%) \\ \hline
1           & 27              & 55.6     & 7.4  \\ \hline
2           & 28              & 53.6     & 28.6  \\ \hline
3           & 23              & 26.1     & 34.8  \\ \hline
4           & 47              & 61.7     & 2.1   \\ \hline
5           & 8               & 0.0      & 0.0   \\ \hline
6           & 38              & 31.6     & 2.6   \\ \hline
7           & 6               & 33.3     & 16.7   \\ \hline
8           & 20              & 15.0     & 0.0     \\ \hline
9           & 13              & 23.1     & 61.5   \\ \hline
10          & 9               & 66.7     & 0.0    \\ \hline
11          & 12              & 16.7     & 25.0    \\ \hline
12          & 11              & 81.8     & 0.0     \\ \hline
\end{tabular}
\label{table:Pred_eval}
\end{table}

\newpage
\section{Conclusions}
The riser in the field experiences complex environmental load conditions. The current profile and its directionality, wave loads and vessel motion induced flow velocities all affect riser responses, which are represented by a 7-parameter space. 

Gaussian mixture models can separate environmental conditions into clusters based on the statistical similarity of the data assisted by domain knowledge. A total of 242 measurement events were investigated and grouped into 12 clusters using only the environmental condition parameters. The number of cluster selected was based on the quality of the separation by the GM algorithm and the data size of the identified clusters. The cluster results were also evaluated manually based on the local acceleration measurements. Both approaches gave consistent results.

Insights on riser responses were obtained by evaluating the characteristic values of environmental load parameters in the identified clusters. Cluster analysis is a valuable tool to reduce the parameter dimensions so that the underlying physics can be better understood.

Riser response prediction using the time domain VIV prediction tool was conducted using the actual 3-directional current and vessel motions, and the comparison with measurements is satisfactory. The potential reasons for prediction deviations were better explained based on the cluster results. This indicates the necessity of performing integrated analysis, in which multiple load processes can be included simultaneously. Adaptive empirical hydrodynamic parameters may also be required when the load process becomes different than in a constant flow condition, and this will be a topic for further studies. 

\section{Acknowledgement}
The authors are grateful for the financial support to the research project PRAI (Prediction Riser-response by Artificial Intelligence, project no.\ 308832) from the Research Council of Norway, Equinor, BP, Subsea7, Kongsberg Maritime and Aker Solutions, and their permission to publish this work. The authors are thankful to Dr.\ Guttorm Grytøyr for his help to provide data and information on the Helland-Hansen riser measurements. 

\newpage
\bibliography{AOR_Ref}
\bibliographystyle{abbrvnat}






\end{document}